\documentclass[12pt]{article}
\usepackage{amsfonts}
\begin{document}
\input amssym.def
\input amssym
\centerline{\bf (p,q)-Sheaves and the Representability Problem }

\centerline {A. I. Panin}

\centerline{ \sl Chemistry Department, St.-Petersburg State University,}
\centerline {University prospect 26, St.-Petersburg 198504, Russia }
\centerline { e-mail: andrej@AP2707.spb.edu }
\bigbreak
{\bf ABSTRACT: }{\small 
General properties of new models of the electronic Fock spaces based on the 
notion of $(p,q)$-sheaves  are studied. Interrelation between simple sheaves
and density operators is established.  Explicit expressions for the transformed
reduced Hamiltonians in terms of the standard creation-annihilation operators
are presented. General scheme of parametrization of $p$-electron states by 
$\varkappa $-electron means $(\varkappa =2,3,\ldots )$ is described and 
studied in detail for the case of sheaves induced by $\varkappa $-electron 
wavefunctions. It is demonstrated that under certain conditions $p$-electron
problem may be reformulated as the eigenvalue problem in $\varkappa$-electron
space equipped with certain $p$-electron metric. Simple numerical examples are 
given to illustrate our approach.  
}

\bigbreak
{\bf Key words: }{\small representability problem; density
operators; CI method; }

{\small electron correlation.}

\bigbreak
\hrule
\bigbreak
{\bf Introduction}

\bigbreak
In present 
work we continue  study of properties of new models of 
$p$-electron sections of
the  Fock spaces introduced in our previous work {\cite {Panin-1}}.
These models are based on the notion of $(p,q)$-sheaves that can be 
considered as $q$-electron representation of $p$-electron states 
$(q\le p)$. Figuratively speaking, the space of all $(p,q)$-sheaves
is a model of $p$-electron space `ready for $q$-electron interactions'.
This means that no contraction is required for calculation of matrix elements
of arbitrary $q$-electron operators. The contraction is replaced by the
summation over $q$-electron functions (germs) constituting the sheaf.

In the second section necessary basic definitions are given.

In the third section general properties of simple $(p,q)$-sheaves 
are studied. Important notion of vector $Z$-cell is introduced.
Explicit characterization of imagies of simple sheaves with respect 
to different compositions of the assembling-disassembling mappings 
is given.

In the fourth section it is shown that there exists a one-to-one 
correspondence (up to arbitrary overall phase prefactor) between simple sheaves
and a certain class of density operators.

In the fifth section the explicit expression for the transformed reduced 
Hamiltonian in terms of the creation-annihilation operators is obtained.

In the sixth section the  notion of $p$-electron metric in 
$\varkappa$-electron space $(\varkappa =1,2,3,\ldots)$ is introduced 
and it is demonstrated that under certain conditions $p$-electron
problem reduces to the eigenvalue problem in $\varkappa$-electron
space equipped with $p$-electron metric.

In the seventh section orbital representation of $(p,q)$-sheaves is
discussed and relevant formulas for metrics and Hamiltonians  are
derived in terms of molecular orbitals.

The eighth section is dedicated to numerical testings.
\bigbreak
\hrule     
\bigbreak

{\bf Basic Definitions.}
\bigbreak

For fixed basis set of $n$ orthonormal spin orbitals the corresponding finite-dimensional 
Fock space ${\cal F}_n$ is  spanned by determinants 
$|R\rangle $ where $R$ runs over all subsets of the spin-orbital index set.
Basis determinants will be  labelled by {\sl subsets}   and  all sign conventions
connected with their  representation as  the Grassman
product of {\sl ordered} spin-orbitals will be included in the definition of the
creation-annihilation operators and in the definiton of specific set-theoretical
operation $\Delta_K, K\subset N$ that was introduced in {\cite{Panin-2}} and studied in detail 
in {\cite{Panin-3}}.Here $N$ is 
the spin-orbital index set.
  
For any $(p+q)$-element subset $Z$ of the index set $N$ let us
denote by the symbol ${\cal F}_{n,q}(Z)$ the following subspace of the 
$q$-electron sector of the Fock space:
 
$${\cal F}_{n,q}(Z)=\bigoplus\limits_{S\subset Z}^{(q)}
{\Bbb C}|S\rangle   
\eqno(1)$$

where $\Bbb C$ is the field of complex numbers.
Let us define the set
$$B_{n,p,q}=\{(Z,S)\subset N\times N:|Z|=p+q \& |S|=q \& S\subset Z\},\eqno(2)$$
and the equivalence relation on this set
$$ (Z,S) \sim (Z',S') \Leftrightarrow Z\backslash S=Z'\backslash S'. \eqno(3)$$
The set of the corresponding  equivalence classes contains  $n\choose p$
elements and in each equivalence class there are $n-p\choose q$
elements.

{\bf Definition.} Family $\{{\psi}_Z\}_{Z\subset N}$ of $q$-electron
functions, ${\psi}_Z\in {\cal F}_{n,q}(Z)$, is called a sheaf of q-electron
germs of $p$-electron wavefunction, or just a $(p,q)$-sheaf, if the mapping
$$ (Z,S) \to (-1)^{|(Z\backslash S)\cap \Delta_Z|}\langle S|{\psi}_Z\rangle,
|Z|=p+q,|S|=q,S\subset Z\eqno(4)$$
is constant on the equivalence classes of the set $B_{n,p,q}$ modulo the
equivalence relation (3). Here $\Delta_Z$ is the  mentioned above 
set-theoretical
operation that is used  to reduce manipulations 
with sign prefactors
in Grassman algebra to pure set-theoretical ones. 

The set of all $(p,q)$-sheaves is denoted as ${\cal S}_{n,p,q}$.

In paper {\cite{Panin-1}} it was demonstrated that there exists a one-to-one 
correspondence between the set of all $(p,q)$-sheaves and vectors of 
the $p$-electron sector of the Fock space. The mapping
$$s_{n,p,q}:\sum\limits_{R\subset N}^{(p)}C_R|R\rangle \to 
{\{\psi}_Z\}_{Z\subset N}\eqno(5)$$
where
$$|\psi_Z\rangle =\sum\limits_{S\subset Z}^{(q)}(-1)^{|(Z\backslash S)
\cap \Delta_Z|}C_{Z\backslash S}|S\rangle. \eqno(6)$$
is called the disassembling mapping and is used to transfer all
relevant structures from the $p$-electron sector of the Fock space on the set 
of all $(p,q)$-sheaves. 
The  {\sl linear structure} is transfered as
$$s_{n,p,q}(\sum\limits_i {\lambda}_i{\Psi}^{(i)})=\left \{
(\sum\limits_i {\lambda}_i{\Psi}^{(i)})_Z \right\}_{Z\subset N}=
\sum\limits_i {\lambda}_i\{ {\psi}^{(i)}_Z\}_{Z\subset N}.\eqno(7)$$
This equality just expresses the fact the due to the {\sl linear character}
of the 
gluing conditions (4) arbitrary linear combination of $(p,q)$-sheaves is 
a $(p,q)$-sheaf. 

The {\sl inner product (Euclidean structure)} is defined as
$$\langle \{{\psi}_Z\}_{Z\subset N}|\{{\phi}_Z\}_{Z\subset N}\rangle
=\frac{1}{{{n-p}\choose q}}
\sum\limits_{Z\subset N}^{(p+q)}\langle {\psi}_Z|{\phi}_Z\rangle \eqno(8)$$
and it is consistent with the inner product of the corresponding $p$-electron
wavefunctions.

The inverse to $s_{n,p,q}$ is called the assembling 
mapping and is used to restore  $p$-electron vectors from 
$(p,q)$-sheafs with the aid of the gluing conditions (4). 

In paper {\cite{Panin-1}} it 
was mentioned that from the very beginning it seems reasonable to associate
with each $p$-electron wavefunction $\Psi$ its sheaf $s_{n,p,q}(\Psi)$ and co-sheaf
$s_{n,n-p,q}(\Bbb I\Psi)$ where $\Bbb I:|R\rangle \to |N\backslash R
\rangle$ is the particle-hole involution. However, since general properties of
sheaves and co-sheaves are  identical, and actual difference appears only on 
the level of evaluation of reduced Hamiltonian matrix elements, we study
properties of sheaves, keeping in mind that the analogous properties 
are valid also for co-sheaves.

The family 
$$\{
{\pi}_{n,p,q}(Z):\{ {\psi}_{Z'}\}_{Z'\subset N} \to {\psi}_Z
\}_{Z\subset N}\eqno(9)$$
involves linear  mappings that perform  projections of the vector 
space  ${\cal S}_{n,p,q}$ on the vector spaces ${\cal F}_{n,q}(Z)$.

Let us suppose that some $q$-electron wavefunction ${\psi }_Z\in {\cal F}_{n,q}
(Z)$ is chosen: 
$${\psi}_Z=\sum\limits_{S\subset  Z}^{(q)}c_S|S\rangle =
\sum\limits_{R\subset  Z}^{(p)}(-1)^{|R\cap {\Delta}_Z|}{\bar C}_R|Z\backslash R\rangle
\eqno(10)$$
where  $R=Z\backslash S$ and
$${\bar C}_R=(-1)^{|R\cap {\Delta}_Z|}c_S\eqno(11)$$
Since ${\cal F}_{n,q}(Z)$ is isomorphic to ${\cal F}_{n,p}(Z)$, we can
easily lift $q$-electron wavefunction to the level $p$ using, for example, 
the linear mapping:

$$u_q^p(Z):
\sum\limits_{R\subset  Z}^{(p)}(-1)^{|R\cap {\Delta}_Z|}{\bar C}_R|Z
\backslash R\rangle \to 
\sum\limits_{R\subset  Z}^{(p)}{\bar C}_R|R\rangle\eqno(12)$$

Subsequent disassembling of $p$-electron wavefunction 
$u_q^p(Z){\psi }_Z$ gives us $(p,q)$-sheaf that may be considered
as generated by $q$-electron wavefunction ${\psi }_Z$. Compositions 
$s_{n,p,q}\circ u_q^p(Z)=j_{n,p,q}(Z)$ constitute 
the family of linear injective mappings
$$j_{n,p,q}(Z):{\cal F}_{n,q}(Z) \to {\cal S}_{n,p,q}\eqno(13)$$  
that are  right inverses (sections) of the projections (9). 

{\bf Definition.}$(p,q)$-sheaf generated by $q$-electron wavefunction  
${\psi }_Z$ is called simple and is denoted as
$\{{\psi}_{ZZ'}\}_{Z'\subset N}$.

It is to be noted that the same simple $(p,q)$-sheaf may be generated 
by  different $q$-electron wavefunctions. 

$(p,q)$-sheaves corresponding to
single determinant $p$-electron wavefunctions are called determinant sheaves.
The characteristic property of determinant sheaves is: simple $(p,q)$-sheaf
is the determinant one if and only if it is 
generated by any of its nonzero germ (see{\cite{Panin-1}}).

It is clear that the assembling and disassembling mappings may be used to
establish isomorphism between operator spaces over ${\cal F}_{n,p}$ and 
over ${\cal S}_{n,p,q}$. Indeed, if $h\in End_{\Bbb C}({\cal F}_{n,p})$ 
then the mapping
$$h\to s_{n,p,q}\circ h\circ s^{-1}_{n,p,q}\eqno(14)$$
is the required isomorphism. For physically relevant operators it is usually
possible to express their matrix elements between two $(p,q)$-sheaves 
as a sum of matrix elements of  certain $q$-electron operators between
germs of these sheaves. 
In our previous paper {\cite{Panin-1}} the electronic Hamiltonian $H$ was handled in such a
way. Namely, due to the specific form of $H$, its matrix element between two 
$(p,q)$-sheaves may be written as 
$$\langle\{ {\phi}_Z\}_{Z\subset N}|H|\{ {\psi}_Z\}_{Z\subset N}\rangle=
\frac{1}{{{n-p}\choose q}}\sum\limits_{Z\subset N}^{(p+q)}
\langle {\phi}_Z|A^{-1}(n,p,q)H_{p\to q}|
{\psi}_Z\rangle,\eqno(15)$$
where 
$$H_{p\to q}=\frac{{p\choose 2}}{{q\choose 2}}\left [
\frac{q-1}{p-1}\sum\limits_{i,j}\langle i|h|j
\rangle a_i^{\dag}a_j+
{1\over 2}\sum\limits_{i,j,k,l}\langle ij|{1\over{r_{12}}}|kl\rangle
a_i^{\dag}a_j^{\dag}a_la_k\right ]\eqno(16)$$
is the so-called reduced Hamiltonian and $A(n,p,q)$ is introduced 
in {\cite{Panin-3}} operator acting on the operator space 
${\cal F}_{n,q}\otimes {\cal F}_{n,q}^*$.
\bigbreak
\hrule
\bigbreak

{\bf Properties of Simple Sheaves.}
\bigbreak

With each function ${\psi}_Z\in {\cal F}_{n,q}(Z)$ written in the form of 
Eq.(10) let us associate two sets:
$$I_p({\psi}_Z)=\{R\subset Z:|R|=p \ \& \ \bar C_R \ne 0\}\eqno(17)$$
$$L({\psi}_Z)=\bigcup\limits_{R\in I_p({\psi}_Z)}R \eqno(18)$$

Let  $\{{\psi}_{ZZ'}\}_{Z'\subset N}$ be some simple $(p,q)$-sheaf 
generated by $q$-electron wavefunction ${\psi}_Z$ .
Then any germ of this sheaf may be written as

$${\psi}_{ZZ'}=\sum\limits_{R'\subset  Z'\cap Z}^{(p)}(-1)^{|R'\cap {\Delta}_{Z'}|}
\bar C_{R'}|Z'\backslash R'\rangle, \eqno(19)$$
where the coefficients $\bar C_{R'}$ are given by Eq.(11).

{\bf Proposition 1.} $q$-electron wavefunction generating simple
$(p,q)$-sheaf  is necessarily a germ of this sheaf.

{\bf Proof.} Follows directly from the relation 
$$ {\pi}_{n,p,q}(Z)\circ j_{n,p,q}(Z)=id_{{\cal F}_{n,q}(Z)}$$
where $id_{{\cal F}_{n,q}(Z)}$ is the identity operator over 
${\cal F}_{n,q}(Z)\blacksquare$ 

{\bf Proposition 2.} $q$-electron functions  ${\psi}_{Z_1}$ and
${\psi}_{Z_2}$ generate the same simple $(p,q)$-sheaf if and only if 

(i) $I_p({\psi}_{Z_1})=I_p({\psi}_{Z_2});$

(ii) For any $R\in I_p({\psi }_{Z_1})$ functions ${\psi}_{Z_1}$ and
${\psi}_{Z_2}$ are subject to the following "gluing" conditions 

$$(-1)^{|R\cap {\Delta }_{Z_1}|}\langle Z_1\backslash R|{\psi }_{Z_1}\rangle =
(-1)^{|R\cap {\Delta }_{Z_2}|}\langle Z_2\backslash R|{\psi }_{Z_2}\rangle .$$

{\bf Proof.} Let us write down $q$-electron functions ${\psi}_{Z_1}$ and
${\psi}_{Z_2}$ in the form of Eq.(10)
$${\psi}_{Z_i} =
\sum\limits_{R\subset  Z_i}^{(p)}(-1)^{|R\cap {\Delta}_{Z_i}|}{\bar C}_R^{(i)}
|Z_i\backslash R\rangle \ (i=1,2)$$
From the equality $\{{\psi}_{Z_1Z'}\}_{Z'\subset N} = 
\{{\psi}_{Z_2Z'}\}_{Z'\subset N}$ it follows that ${\psi}_{Z_1}={\psi}_{Z_2Z_1}$
and ${\psi}_{Z_2}={\psi}_{Z_1Z_2}$. These equalities together with Eq.(19) 
imply that $\bar C_R^{(1)}=\bar C_R^{(2)}$, and, consequently, condition (i) and (ii) are
fulfilled.    

Now let us suppose that for given functions ${\psi}_{Z_1}$ and
${\psi}_{Z_2}$ the conditions (i) and (ii) hold true. Then
from (i) it follows that for arbitrary $Z'\subset N$
$${\psi}_{Z_iZ'}=\sum\limits_{R\subset  Z'\cap Z_1 \cap Z_2}^{(p)}(-1)^{|R\cap {\Delta}_{Z'}|}
\bar C_{R}^{(i)}|Z'\backslash R\rangle, $$
where i=1,2. Use of conditions (ii) immediately leads to the equality
  
$\{{\psi}_{Z_1Z'}\}_{Z'\subset N} = \{{\psi}_{Z_2Z'}\}_{Z'\subset N}
\ \blacksquare$

{\bf Proposition 3.} The number of $q$-germs generating simple $(p,q)$-
sheaf $\{{\psi}_{ZZ'}\}_{Z'\subset N}$ is equal to
$$ {n-|L({\psi }_Z)|}\choose {p+q-|L({\psi }_Z)|}$$.
  
{\bf Proof.} By definition, $L({\psi }_Z)\subset Z$. For any $Z_1\supset
L({\psi }_Z)$ the germs ${\psi }_Z$ and ${\psi}_{ZZ_1}$ obviously
satisfy the conditions of proposition 2 $\blacksquare $
            
{\bf Corollary 1.} For any determinant sheaf $|L({\psi }_Z)|=p$ and it is 
generated by any of its ${n-p}\choose q$ germs.

{\bf Corollary 2.} Non-determinant $(p,1)$-sheaf has only one generator.

{\bf Proof.} For non-determinant sheaves $|L({\psi }_Z)|$ should necessarily 
be equal to $p+1$ $\blacksquare $
  
{\bf Definition.} $q$-electron functions  ${\psi}_{Z_1}$ and
${\psi}_{Z_2}$ are called $p$-equivalent 
(${\psi}_{Z_1}\stackrel{p}{\sim }{\psi}_{Z_2}$) 
if they generate the same simple $(p,q)$-sheaf. 

{\bf Proposition 4.}  Let $\{{\psi}_{ZZ'}\}_{Z'\subset N}$ be simple
$(p,q)$-sheaf generated by $q$-electron wavefunction ${\psi}_Z$. Then
any $q$-electron function from ${\cal F}_{n,q}(Z)$ is orthogonal to all
${\psi}_{ZZ'}$ with $Z'\ne Z$.

{\bf Proof.} Expansion of arbitrary $q$-electron function from 
${\cal F}_{n,q}(Z)$ (see Eq.(10)) involves
$q$-electron determinants $|Z\backslash R\rangle $ where $R\subset Z$. From 
Eq.(19) it follows that each $q$-germ ${\psi}_{ZZ'}$ of the sheaf under 
consideration is expanded via $q$-electron determinants 
$|Z'\backslash R'\rangle$. It is easy to see that $Z'\ne Z$ and 
$R'\subset Z'\cap Z$ implies $Z'\backslash R' \not \subset Z$\ $\blacksquare$
 
{\bf Corollary.} $q$-germs generating simple $(p,q)$-sheaf are mutually
orthogonal.

Let us consider a family of $k$ simple $(p,q)$-sheaves 
$\{{\psi}_{Z'_iZ'}\}_{Z'\subset N}$, $i=1,2,\ldots,k$ and sort out this family
in the following manner:

Select some subset $Z_1$ from the family $Z'_1,Z'_2,\ldots,Z'_k$ and 
collect all $(p,q)$-sheaves such that 
${\psi}_{Z'_iZ_1}\stackrel{p}{\sim }{\psi}_{Z'_i}$. The resulting subfamily
will be denoted as $\{{\psi}_{Z_1Z'}^{(i_1)}\}_{Z'\subset N}, i_1=1,2,\ldots,k_1$;

From the rest of the initial family select subset $Z_2$ and  
collect all $(p,q)$-sheaves such that 
${\psi}_{Z'_iZ_2}\stackrel{p}{\sim }{\psi}_{Z'_i}$. The resulting subfamily
will be denoted as $\{{\psi}_{Z_2Z'}^{(i_2)}\}_{Z'\subset N}, i_2=1,2,\ldots,k_2$.

Repeating the last step till the initial family is exausted, we arrive finally
at the family $\{{\psi}_{Z_lZ'}^{(i_l)}\}_{Z'\subset N},i_l=1,2,
\ldots,k_l,  l=1,2,\ldots,$. By construction, each subfamily of this family 
corresponding to some fixed index l involves simple $(p,q)$-sheaves 
generated by
$q$-germs from the vector space ${\cal F}_{n,q}(Z_l)$. As a result, the 
problem of orthogonalization of simple sheaves within this subfamily seems to
be  not very complicated, especially if one takes into account the following 
assertion.

{\bf Proposition 5.} Let  $\{{\psi}_{ZZ'}^{(1)}\}_{Z'\subset N}$ and
$\{{\psi}_{ZZ'}^{(2)}\}_{Z'\subset N}$  be  simple $(p,q)$-sheaves generated
by $q$-electron functions ${\psi}_Z^{(1)}, {\psi}_Z^{(2)}\in 
{\cal F}_{n,q}(Z)$. Then
$$\langle \{{\psi}_{ZZ'}^{(1)}\}_{Z'\subset N}|\{{\psi}_{ZZ'}^{(2)}\}_{Z'\subset N}
\rangle = \langle  {\psi}_Z^{(1)}|{\psi}_Z^{(2)} \rangle. \eqno(20)$$

{\bf Proof.} Direct calculation based on the definition (8) of the inner 
product leads readily to the equality required$\ \blacksquare$

The question when linear combination of simple sheaves is simple is of a
certain interest in connection with the problem of selection of linearly
independent simple sheaves. The answer to this question is given by the
following assertion.

{\bf Proposition 6.} Linear combination
$$\sum\limits_{i=1}^k {\lambda}_i\{{\psi}_{Z_iZ'}\}_{Z'\subset N}$$
of simple $(p,q)$-sheaves is simple if and only if there exists
$(p+q)$-element subset $Z\subset N$ such that 
$$R\not\subset Z \Rightarrow \sum\limits_{i\in I_R^{Z_1\ldots Z_k}}
{\lambda}_i\bar C_R^{(i)}=0,$$
where
$$I_R^{Z_1\ldots Z_k}=\{1\le i\le k:Z_i\supset R\}.$$

{\bf Proof.} Arbitrary linear combination of simple sheaves may be written 
in the form
$$\sum\limits_{i=1}^k{\lambda}_i\{{\psi}_{Z_iZ'}\}_{Z'\subset N}=
\left \{ \sum\limits_{R'\subset Z'}^{(p)}|Z'\backslash R'\rangle 
(-1)^{|R'\cap {\Delta}_{Z'}|}\left (\sum\limits_{i\in I_{R'}^{Z_1\ldots Z_k}}
{\lambda}_i\bar C_{R'}^{(i)}\right )\right \}_{Z'\subset N}.$$
If we suppose that this sheaf is simple then there should exist index set 
$Z\subset N$ such that the $q$-germ 
$$\sum\limits_{R\subset Z}^{(p)}|Z\backslash R\rangle 
(-1)^{|R\cap {\Delta}_Z|}\left (\sum\limits_{i\in I_R^{Z_1\ldots Z_k}}
{\lambda}_i\bar C_R^{(i)}\right )$$
generates this sheaf. It can, however, be true if and only if the 
coefficients  by the basis determinants $|Z'\backslash R'\rangle$ 
in expansion of $q$-germs are equal to zero for all $R'\not\subset Z\ 
\blacksquare$

{\bf Corollary.} Arbitrary linear combination of determinant $(p,q)$-sheaves
generated by $q$-germs
$${\psi}_{Z_i}=(-1)^{|R_i\cap {\Delta}_{Z_i}|}\bar C_{R_i}^{(i)} 
|Z_i\backslash R_i \rangle $$
is simple if and only if there exists index set $Z\subset N$ such that 
$Z\supset \bigcup\limits _{i=1}^kR_i$, or, in other words, 
$|\bigcup\limits _{i=1}^kR_i|\le p+q$.

{\bf Proof.} Without loss of generality we may assume that the coefficients 
${\lambda }_i$ of linear combination of determinant sheaves are non-zero.
Let us suppose that for any subset $Z\subset N$ there exists $R_j\not\subset Z$.
In this case 
$$\sum\limits_{i\in I_{R_j}^{Z_1\ldots Z_k}}
{\lambda}_i\bar C_{R_j}^{(i)}={\lambda}_j\bar C^{(j)}_{R_j}\ne 0$$
and, consequently, the linear combination under consideration can not be simple
$\blacksquare$

{\bf Definition.} The image of ${\cal F}_{n,q}(Z)$ with respect to the 
monomorphism $j_{n,p,q}(Z)$ is called the vector $Z$-cell of the vector
space of all $(p,q)$-sheaves and is denoted as $J_{n,p,q}(Z)$.

It is pertinent to note that $Z$-cells depend on  the MSO basis chosen.

Each vector $Z$-cell is a ${p+q}\choose p$-dimensional subspace of 
${\cal S}_{n,p,q}$. The set of all simple $(p,q)$-sheaves in 
${\cal S}_{n,p,q}$ is a union of its vector $Z$-cells:
$$J_{n,p,q}=\bigcup\limits_{Z\subset N}^{(p+q)}J_{n,p,q}(Z).\eqno(21)$$
It is easy to show that the set $J_{n,p,q}$ generates the vector space of
all $(p,q)$-sheaves (see {\cite{Panin-1}}).

The set of simple $(p,q)$-sheaves $\{{\psi}_{ZZ'}\}_{Z'\subset N}\in
J_{n,p,q}(Z)$ such that 
$$|L({\psi}_Z)|\le k $$
where $k=p+q,p+q-1,\ldots,p$, will be denoted as $J_{n,p,q}^{(k)}(Z)$.
It is clear that each $Z$-cell  admits the following filtration
$$J_{n,p,q}(Z)=J_{n,p,q}^{(p+q)}(Z)\supset J_{n,p,q}^{(p+q-1)}(Z)\supset
\ldots \supset J_{n,p,q}^{(p)}(Z)\eqno(22)$$
{\sl Topological } dimension of $J_{n,p,q}^{(k)}(Z)$ is equal to $k\choose p$.
Thus, each $Z$-cell possesses $k$-faces (borders) where simple $(p,q)$-sheaves from
different cells are situated. It is likely that filtration (21) may serve as a 
starting point for further more advanced  analysis of geometry of the set 
$J_{n,p,q}$.   

Now let us consider the connection between spaces ${\cal S}_{n,p,q}$ with 
different $q$. There exist obvious isomorphisms
$${\cal S}_{n,p,q}\stackrel {s^{-1}_{n,p,q}}{\longrightarrow}{\cal F}_{n,p}
\stackrel {s_{n,p,q\pm 1}}{\longrightarrow}{\cal S}_{n,p,q\pm 1}.\eqno(23)$$

{\bf Proposition 7.} The image of any simple $(p,q)$-sheaf from $J_{n,p,q}$
with respect to the isomorphism $s_{n,p,q+1}\circ s^{-1}_{n,p,q}$ is simple
$(p,q+1)$-sheaf, and for $k=p+q, p+q-1,\ldots,p$  
$$s_{n,p,q+1}\circ s^{-1}_{n,p,q}\left (J^{(k)}_{n,p,q}\right )
\subset J^{(k)}_{n,p,q+1},\eqno(24)$$
that is the image of $J_{n,p,q}$ belongs to the border of $J_{n,p,q+1}.$

{\bf Proof.} Let ${\psi}_Z\in {\cal F}_{n,q}(Z)$ generate simple $(p,q)$-
sheaf $\{{\psi }_{ZZ'}\}_{Z'\subset N}$ and $u_q^p(Z){\psi}_Z=
\sum\limits_{R\subset Z}^{(p)}\bar C_R|R\rangle ={\Psi}_Z$. For {\sl any}
index $i\in N\backslash Z$ the $p$-electron  wavefunction
$${\Psi}_{Z\cup \{i\}}=\sum\limits_{{R\subset Z\cup i}\atop {(R\ni i)}}^{(p)}
\bar C_R|R\rangle =
{\Psi}_Z$$
can be disassembled to give simple $(p,q+1)$-sheaf with germs
$$ {\psi}_{Z\cup \{i\}Z'}=\sum\limits_{R\subset Z'\cap Z}^{(p)}
(-1)^{|R\cap {\Delta}_{Z'}|}\bar C_{R}|Z'\backslash R\rangle$$
where $|Z'|=p+q+1$. It is clear as well that the choice of concrete index $i$ is 
irrelevant. Indeed, for any index $j\in N\backslash Z$, $(q+1)$-germ 
${\psi}_{Z\cup \{i\}Z\cup \{j\}}$ is $p$-equivalent to ${\psi}_{Z\cup \{i\}}$
$\blacksquare$

Lowering of simple $(p,q)$-sheaf to the level $q-1$  leads, in general, to
linear combinations of simple $(p,q-1)$-sheaves.

{\bf Proposition 8.} Let $\{{\psi}_{ZZ'}\}_{Z'\subset N}$ be simple 
$(p,q)$-sheaf generated by $q$-germ ${\psi }_Z$. Then
$$s_{n,p,q-1}\circ s^{-1}_{n,p,q}\left (\{{\psi}_{ZZ'}\}_{Z'\subset N}\right )=
\frac{1}{q}\sum\limits_{k\in Z}\{{\psi}_{Z\backslash \{k\} Z''}\}_{Z''\subset N}.
\eqno(25)$$

{\bf Proof.} Disassembling of $p$-electron function ${\Psi}_Z=
u^p_q(Z){\psi}_Z, |Z|=p+q$, gives $(p,q-1)$-sheaf with germs (see Eq.(6))
$${\psi}_{Z''}=\sum\limits_{R\subset Z''\cap Z}^{(p)}(-1)^{|R\cap {\Delta}_{Z''}|}
\bar C_R|Z''\backslash R\rangle,\eqno(26)$$
where $|Z''|=p+q-1$.
 
On the other hand, let us consider the sum 
$$\sum\limits_{k\in Z}\{{\psi}_{Z\backslash \{k\}Z''}\}_{Z''\subset N}=
\left \{ \sum\limits_{k\in Z}{\psi}_{Z\backslash \{k\}Z''} \right \}_{Z''\subset N},
\eqno(27)$$
where simple $(p,q-1)$-sheaf $\{{\psi}_{Z\backslash \{k\}Z''}\}_{Z''\subset N}$
involves $(q-1)$-germs
$${\psi}_{Z\backslash \{k\}Z''}=
\sum\limits_{R\subset [Z''\cap (Z\backslash \{k\})]}^{(p)}
(-1)^{|R\cap {\Delta}_{Z''}|}
\bar C_R|Z''\backslash R\rangle.$$

Thus, $(p,q-1)$-sheaf on the right-hand side of Eq.(27) involves $(q-1)$-germs
$$\sum\limits_{k\in Z} 
\sum\limits_{R\subset [Z''\cap (Z\backslash \{k\})]}^{(p)}
(-1)^{|R\cap {\Delta}_{Z''}|}
\bar C_R|Z''\backslash R\rangle.$$
The contribution to this sum from  $k\not\in Z''\cap Z$ is equal to
$(p+q-|Z''\cap Z|){\psi}_{Z''}$ where ${\psi}_{Z''}$ is given by Eq.(26).
On the other hand, for each fixed $R\subset Z''\cap (Z\backslash \{k\})$ there
exist $|Z''\cap Z|-p$ different $k\in Z''\cap Z$. As a result,  

$$\sum\limits_{k\in Z}\{{\psi}_{Z\backslash \{k\}Z''}\}_{Z''\subset N}=
q\left \{{\psi}_{Z''}\right \}_{Z''\subset N} \blacksquare$$

It is to be noted that if $L({\psi}_Z)$ is {\sl a proper subset} of Z then
for any $k\in Z\backslash L({\psi}_Z)$ (the choice of $k$ is irrelevant), 
$s_{n,p,q-1}\left ({\Psi}_Z\right )=s_{n,p,q-1}\left ({\Psi}_{Z\backslash \{k\}}\right )$ 
is simple $(p,q-1)$-sheaf.

{\bf Corollary.}  For any admissible $q$ and any $\varkappa < q
$
$$s_{n,p,\varkappa}\circ s^{-1}_{n,p,q}\left (\{{\psi}_{ZZ'}\}_{Z'\subset N}\right )=
\frac{\varkappa !}{q!}\sum\limits_{K\subset Z}^{(q-\varkappa)}\{{\psi}_{Z\backslash K Z''
}\}_{Z''\subset N}.
\eqno(28)$$
Very important is the following statement that  is contrary in a certain sense 
to the Proposition 8.

{\bf Proposition 9.} If $Z\subset N, |Z|=p+q$, and for each $(q-\varkappa)$-
element subset $K\subset Z$
$${\psi}_{Z\backslash K}=\sum\limits_{R\subset Z\backslash K}^{(p)}
(-1)^{|R\cap {\Delta}_{Z\backslash K}|}\bar C_R^{(K)}|Z\backslash K \backslash R
\rangle ,\eqno (29)$$
then the image of the sum 
$$\sum\limits_{K\subset Z}^{(q-\varkappa)}\{{\psi}_{Z\backslash K Z'
}\}_{Z'\subset N}$$
with respect to the assembling mapping $s^{-1}_{n,p,\varkappa}$ is $p$-electron
wavefunction 
$${\Psi}_Z=\sum\limits_{R\subset Z}^{(p)}
\left (\sum\limits_{K\subset Z\backslash R}^{(q-\varkappa)}\bar C_R^{(K)}\right )
|R\rangle ,\eqno(30)$$
and its disassembling with the aid of $s_{n,p,q}$ gives simple $(p,q)$-sheaf
generated by the $q$-germ
$$\sum\limits_{R\subset Z}^{(p)}(-1)^{|R\cap {\Delta}_Z|} 
\left (\sum\limits_{K\subset Z\backslash R}^{(q-\varkappa)}\bar C_R^{(K)}\right )
|Z\backslash R\rangle .\eqno(31)$$
{\bf Proof.} By direct calculation.

This Proposition can be used to parametrize subsets of the sets $J_{n,p,q}(Z)$
using $\varkappa$-electron means ($\varkappa <q$). 
\bigbreak
\hrule
\bigbreak

{\bf Simple Sheaves and Density Operators}

\bigbreak
\bigbreak

{\bf Definition.} The quadratic mapping 
$d_q:{\cal S}_{n,p,q}\to {\cal F}_{n,q}\otimes {\cal F}_{n,q}^*$
defined by
$$d_q\left (\{{\psi}_Z\}_{Z\subset N}\right )=
\frac{1}{{{n-p}\choose q}}\sum\limits_{Z\subset N}^{(p+q)}|{\psi}_Z\rangle
\langle {\psi}_Z|\eqno(32)$$
is called the density mapping of order $q$. The value of this mapping at some
$(p,q)$-sheaf is called the $q$-density operator associated with this sheaf. 

The problem of description of the set $d_q\left ({\cal S}_{n,p,q}\right )$
is called the pure representability problem and in slightly different form
was formulated in the very first papers by Coleman {\cite {Coleman-1,
Coleman-2,Coleman-3,Coleman-4}}. By
an abuse of notation the same symbol $d_q$ will be used for 
the density mapping defined by Eq.(32) and for elements of the set
$d_q\left ({\cal S}_{n,p,q}\right )$

Let us consider density operators corresponding to simple sheaves.  
From
 Eq.(19) it follows that 
$$d_q\left (\{{\psi}_{ZZ'}\}_{Z'\subset N}\right )$$
$$=\frac{1}{{{n-p}\choose q}}\left [ |{\psi}_Z\rangle\langle {\psi}_Z|+
\sum\limits_{Z'\subset N\atop {\left (p<|Z'\cap Z|<p+q\right )}}
^{(p+q)}|{\psi}_{ZZ'}\rangle
\langle {\psi}_{ZZ'}|+ ||{\psi}_Z||^2P^{(q)}_{N\backslash Z}\right ],\eqno(33)$$
where 
$$P^{(q)}_X=\sum\limits_{S\subset X}^{(q)}|S\rangle\langle S| \eqno(34)$$
is the projector associated with the index set $X\subset N$.
 Note that the
representation (33) of the density operator associated with some given simple 
sheaf is not unique (see Proposition 3).

Complete characterization of $q$-density operators corresponding to simple
$(p,q)$-sheaves is given by the assertion that readily follows from Eq.(33)
and Proposition 4.

{\bf Proposition 10.}  $q$-electron operator $d_q$ is associated with simple 
$(p,q)$-sheaf if and only if there exists $(p+q)$-element subset $Z\subset N$
such that

(i) the operator $P^{(q)}_Zd_qP^{(q)}_Z$  
corresponds
to some (not necessarily normalized) pure $q$-electron state, that is 
$P^{(q)}_Zd_qP^{(q)}_Z=|{\psi}_Z\rangle\langle {\psi}_Z|;$

(ii) $\left (I-P^{(q)}_Z\right )d_q
\left (I-P^{(q)}_Z\right )=
\sum\limits_{{Z'\subset N}\atop {\left (p\le |Z'\cap Z|< p+q\right )}}^{(p+q)}
|{\psi}_{ZZ'}\rangle \langle {\psi}_{ZZ'}|$,

where wavefunctions ${\psi}_{ZZ'}$ are {\it uniquely} determined by the eigenfunction
${\psi}_Z$ of the operator $P^{(q)}_Zd_qP^{(q)}_Z$ corresponding to its
non-zero
 eigenvalue.

Thus, for any fixed $(p+q)$-element subset $Z\subset N$ there exists one-to-one
correspondence (up to arbitrary overall phase prefactor) between elements from 
the vector $Z$-cell $J_{n,p,q}(Z)$ and  representable   density
operators of the form of Eq.(33). Reservation `up to arbitrary overall phase 
prefactor' may be removed by turning to the quotient of $J_{n,p,q}(Z)$ obtained 
by identification of sheaves that differ by overall phase prefactor, or
to the projective $Z$-cells if only normalized sheaves are of interest. 
We, however, prefer to use vector
$Z$-cells saying  that simple sheaves parametrize density
operators of the form of Eq.(33) even if  this statement  is not quite correct.    

The simplest density mapping corresponds to the case $q=1$ and its value at some 
simple $(p,1)$-sheaf is equal to
$$d_1(\{{\psi}_{ZZ'}\}_{Z'\subset N})=\frac{1}{n-p}\left [|{\psi}_Z\rangle
\langle {\psi}_Z|+P^{(1)}_{N\backslash Z} \right ],\eqno(35)$$
where $|Z|=p+1$.
The kernal of the operator on the right-hand side of this equation is 
$p$-dimensional subspace of occupied 
one-electron particle states (occupied molecular spin orbitals).

Using proposition 7 and Eq.(33), it is possible to lift one-density operators 
from $d_1(J_{n,p,1})$ to the level $q=2$:
$$d_2(\{{\psi}_{Z\cup \{i\}Z'}\}_{Z'\subset N})=\frac{1}{{{n-p}\choose 2}}
\left [\vphantom{\sum\limits_{Z'\subset N \atop {\left (|Z'\cap Z|=p+1 \right )}}^{(p+2)}} 
|{\psi}_{Z\cup \{i\}}\rangle \langle {\psi}_{Z\cup \{i\}}| \right .$$
$$\left . +\sum\limits_{Z'\subset N \atop {\left (|Z'\cap Z|=p+1 \right )}}^{(p+2)}
|{\psi}_{Z\cup \{i\}Z'}\rangle \langle {\psi}_{Z\cup \{i\}Z'}| + 
P^{(2)}_{N\backslash (Z\cup \{i\})}\right ].\eqno(36)$$
Non-zero contributions to the sum on the right-hand side of Eq.(36) may 
appear in two cases:
$Z'=Z\cup \{j\}, j\in N\backslash Z, j\ne i$ and 
$Z'=Z\backslash k\cup \{i,j\}, k\in Z, i,j\in N\backslash Z$. Analysis of 
these cases leads to the following final expression:
$$d_2(\{{\psi}_{Z\cup \{i\}Z'}\}_{Z'\subset N})=\frac{1}{{{n-p}\choose 2}}
\left [\sum\limits_{k,k'\in Z}c_kc_{k'}^*
\sum\limits_{j\in N\backslash Z} 
(-1)^{\epsilon}|\{j,k\}\rangle \langle \{j,k'\}| +
P^{(2)}_{N\backslash Z}\right ]\eqno(37)
$$
where 
$$
\epsilon = |\left [(Z\backslash \{k\})\Delta (Z\backslash \{k'\})\right ]
\cap {\Delta}_{\{j\}}|=\cases { 
0, &if $k=k'$ or $k,k'>j$;\cr
1, &if $k\le j\ \& \ k'>j$ or $k'\le j\ \& \ k>j$;\cr
2, &if $k,k'\le j$.}
$$ 
By the symbol 
$\{j,k\}$ on the right-hand side of Eq.(37) non-ordered pairs 
of spin-orbital indices (two-element subsets) are denoted. It is pertinent
to emphasize once again that the right-hand side of Eq.(37) does not depend 
on the choice of index $i\in N\backslash Z$ that appears explicitely on the
left-hand side of this equality (see Proposition 7).

It is possible, following Coleman, to replace the requirement of pure 
representability by more simple requirement of representability by an 
ensemble of pure states. Instead of the quadratic $q$-density mapping (32)
let us introduce the  mapping 
$$d_q:{\cal S}_{n,p,q}\otimes {\cal S}_{n,p,q}^* \to
{\cal F}_{n,q}\otimes {\cal F}_{n,q}^*$$
that is defined on generators 
$|\{{\psi}_Z\}_{Z\subset N}\rangle \langle \{{\phi}_Z\}_{Z\subset N}|$ as
$$d_q\left (|\{{\psi}_Z\}_{Z\subset N}\rangle \langle \{{\phi}_Z\}_{Z\subset N}|\right )=
\frac{1}{{{n-p}\choose q}}\sum\limits_{Z\subset N}^{(p+q)}|{\psi}_Z\rangle
\langle {\phi}_Z|\eqno(38)$$
and is continued to ${\cal S}_{n,p,q}\otimes {\cal S}_{n,p,q}^*$ by linearity. 

The set of all  finite convex combinations  of the type
$$\sum\limits_i {\lambda}_i |\{{\psi}_Z^{(i)}\}_{Z\subset N}\rangle 
\langle \{{\psi}_Z^{(i)}\}_{Z\subset N}|$$
where ${\lambda}_i\ge 0$, $\sum\limits_i {\lambda}_i=1$, and 
$||\{{\psi}_Z^{(i)}\}_{Z\subset N}||=1$ will be 
denoted as $Ens({\cal S}_{n,p,q})$. Its image with respect to $d_q$ is
the set of all ensemble representable $q$-density operators and the problem of 
its analytic description is known as the ensemble representability problem.
For $q=1$ its {\sl constructive} solution was found by Coleman 
{\cite{Coleman-1}}. 
\bigbreak
\hrule
\bigbreak

{\bf Explicit Expressions for the Transformed Reduced Hamiltonians}

\bigbreak
\bigbreak

\bigbreak
From Eq.(15) it follows that in the vector space of all $(p,q)$-sheaves the usual 
p-electron Hamiltonian should be replaced by the reduced Hamiltonian 
transformed with the aid of the automorphism $A^{-1}(n,p,q)$ introduced in 
{\cite{Panin-3}}.
For $q=2$ it is not difficult to perform direct numerical transformation 
to construct the required operator. For $q>2$, however, such an approach may 
appear to be complicated because the transforming operator is of huge dimension.  
Fortunately, it is possible to obtain simple explicit expression for 
$A^{-1}(n,p,q)H_{p\to q}$. We start with more general task of calculation of   
$A(n,p,q)$ action on products of the creation-annihilation operators.

Operator $A(n,p,q)$ was defined in {\cite{Panin-3}} via its matrix representation with respect 
to specially selected basis closely related to the basis of determinant
generators:
$$A(n,p,q)e^{IJ}_K=
(-1)^s \frac{{p\choose q}}{{p-s\choose q-s}{n-p\choose q}}
\sum\limits_{K'\subset N\backslash (I\cup J)}^{(q-s)}(-1)^{|K\cap K'|}
\frac{{p+|K\cap K'|-q-1\choose {|K\cap K'|}}}{{q-s\choose |K\cap K'|}}
e^{IJ}_{K'}\eqno(39)$$
where
$e_K^{IJ}=(-1)^{|(I\cup J)\cap \Delta_K|}|I\cup K\rangle \langle J\cup K|$
and $I\cap J=\emptyset , s=|I|=|J|$.
 
Therefore, as the first step, it is necessary to expand the restrictions of 
products of the creation-annihilation operators on $q$-electron sector 
of the Fock space via the basis operators $e^{IJ}_K$. Using
technique of manipulations with phase prefactors developed in {\cite{Panin-3}}, it is easy
to  derive the following formulas:
$$a^{\dag }_ia_i=\sum\limits_{K\subset N\backslash \{i\}}^{(q-1)}
e^{\emptyset \emptyset }_{K\cup \{i\}},\eqno(40a)$$
$$a^{\dag }_ia_j=\sum\limits_{K\subset N\backslash \{i,j\}}^{(q-1)}
e^{\{i\}\{j\} }_K,\qquad  i\ne j, \eqno(40b)$$
$$a^{\dag }_ia^{\dag }_ja_ja_i=\sum\limits_{K\subset N\backslash \{i,j\}}^{(q-2)}
e^{\emptyset \emptyset }_{K\cup \{i,j\}},\qquad i\ne j,  \eqno(40c)$$
$$a^{\dag }_ia^{\dag }_ja_la_i=\sum\limits_{K\subset N\backslash \{i,j,l\}}^{(q-2)}
e^{\{j\}\{l\}}_{K\cup \{i\}}, \qquad i\ne j\ne l, \eqno(40d)$$
$$a^{\dag }_ia^{\dag }_ja_la_k=\sum\limits_{K\subset N\backslash \{i,j,k,l\}}^{(q-2)}
(-1)^{|\{j\}\cap {\Delta}_{\{i\}}|+|\{l\}\cap {\Delta}_{\{k\}}|}
e^{\{i,j\}\{k,l\}}_K,\  i\ne j\ne k \ne l. \eqno(40e)$$

The second step is straightforward. One should apply the operator $A(n,p,q)$ 
to the right-hand sides of Eqs.(40a)-(40e). Simple but somewhat tiresome 
set-theoretical and combinatorial manipulations with extensive use of two 
classic relations (see, e.g.,{\cite{Knuth}})
$$\sum\limits_k (-1)^k{{r-k}\choose m}{s\choose k}={{r-s}\choose {r-m}}$$
and
$$ {{-r}\choose k}=(-1)^k{{r+k-1}\choose k}$$
lead to the following equalities:
$$A(n,p,q)a^{\dag }_ia_j=-\frac{p}{n-p}a^{\dag }_ia_j+{\delta}_{ij}
\frac{q}{n-p}I_q,\eqno(41a)$$
$$A(n,p,q)a^{\dag }_ia^{\dag }_ja_la_k=\frac{{q\choose 2}}{{{n-p}\choose 2}}
\left [
\frac{{p\choose 2}}{{q\choose 2}}a^{\dag }_ia^{\dag }_ja_la_k-
\frac{p}{q} 
\left ({\delta}_{ik}a^{\dagger}_ja_l+{\delta}_{jl}
a^{\dagger}_ia_k 
\right )\right .$$ 
$$\left . +\frac{p}{q} 
\left ({\delta}_{il}a^{\dagger}_ja_k+{\delta}_{jk}
a^{\dagger}_ia_l 
\right )+
\left ( {\delta}_{ik}{\delta}_{jl}-{\delta}_{il}{\delta}_{jk}\right )
I_q \right ]\eqno(41b)$$
where $I_q$ is the $q$-electron identity operator. 
Now, taking into account the equality $A^{-1}(n,p,q)=A(n,n-p,q)$ (see 
{\cite{Panin-3}}), 
we can write down
the explicit expression for the transformed reduced Hamiltonian:

$$A^{-1}(n,p,q)H_{p\to q}=
\frac{1}{2}\left [Tr(h+F_N)\right ]I_q -\frac{{{n-p}\choose 1}}{{q\choose 1}}
\sum\limits_{i,j}\langle i|F_N|j\rangle a_i^{\dagger}a_j$$
$$ +\frac{1}{2}\frac{{{n-p}\choose 2}}{{q\choose 2}}
\sum\limits_{i,j,k,l}\langle ij|kl\rangle
a_i^{\dag}a_j^{\dag}a_la_k,\eqno(42)$$
where the Fock operator associated with some MSO index set $X\subset N$ is 
defined as 
$$F_X=h+\sum\limits_{k \in X}(J_k-K_k).\eqno(43)$$
The first
term on the right-hand side of Eq.(42) is the Hartre-Fock (HF) energy of 
the $n$-electron 
state $|N\rangle$. It is pertinent to emphasize that the operator (42) 
should be applied to wavefunctions from the $q$-electron sector of the 
Fock space. Note also that with the aid of Eqs.(41a-41e) it is possible to 
expand any physically relevant operator via {\sl orthonormal} basis of 
generators $e^{IJ}_K$ thus reducing the matrix element evaluation problem
to the calculation of standard scalar products.

The energy of the vector $Z$-cell in fixed MSO basis for $q>1$ is obviously 
defined as 
$$E(J_{n,p,q}(Z))=\min _{||{\psi}_Z||=1} Tr\left [A^{-1}(n,p,q)H_{p\to q}
d_q(\{ {\psi}_{ZZ'}\}_{Z'\subset N})\right ].\eqno(44)$$
For $q=1$ it is necessary  to lift one-density operator to the level
$q=2$ and then  define the energy of $Z$-cell $J_{n,p,1}(Z)$ as 
$$E(J_{n,p,1}(Z))=\min _{||{\psi}_Z||=1} Tr\left [A^{-1}(n,p,2)H_{p\to 2}
d_2(\{ {\psi}_{Z\cup \{i\}Z'}\}_{Z'\subset N})\right ],\eqno(45)$$
where $i$ is an arbitrary index from $N\backslash Z$.
\bigbreak
\hrule
\bigbreak

{\bf Parametrizations and Induced Sheaves }
\bigbreak
\bigbreak

Arbitrary vector $Z$-cell $J_{n,p,q}(Z)$ is a subspace of the vector space 
${\cal S}_{n,p,q}$ of the dimension ${{p+q}\choose p}$. Its image with respect to
the isomorphism $s_{n,p,\varkappa}\circ s^{-1}_{n,p,q}$ is a sum of 
vector $Z\backslash K$-cells
$$s_{n,p,\varkappa}\circ s^{-1}_{n,p,q}\left (J_{n,p,q}(Z)\right )=
\sum\limits_{K\subset Z}^{(q-\varkappa)}J_{n,p,\varkappa}(Z\backslash K).
\eqno(46)$$
(see Proposition 8) being a subspace of the vector space ${\cal S}_{n,p,\varkappa}$
of the same dimension ${{p+q}\choose p}$. From Proposition 9 it follows that
different choice of $\varkappa$-electron functions of the type of Eq.(29) 
can be used to parametrize different subsets of the vector $Z$-cell 
$J_{n,p,q}(Z)$. General scheme of such  parametrizations may be 
described as follows:

{\bf 1.} Select a certain family of $\varkappa$-electron functions  
$${\psi}_{Z\backslash K}=\sum\limits_{S\subset Z\backslash K}^{(\varkappa)}
c_S^{(K)}|S\rangle \eqno(47)$$
with a reasonable number of free parameters;

{\bf 2.} To each  selected nonzero $\varkappa$-electron function put into
correspondence simple $(p,\varkappa)$-sheaf generated by this function;

{\bf 3.} Construct $(p,\varkappa)$-sheaf 
$$\{{\psi}_Z'\}_{Z'\subset N}=\sum\limits_{K\subset Z}^{(q-\varkappa)}
\{{\psi}_{Z\backslash KZ'}\}_{Z'\subset N},\eqno(48)$$
that, due to Proposition 9, corresponds to a certain simple $(p,q)$-sheaf
from  $Z$-cell $J_{n,p,q}(Z)$. 

For example, functions
$${\psi}_{Z\backslash K}=\cases { (-1)^{|R\cap {\Delta}_{Z\backslash K}|}
\bar C_R |Z\backslash R\backslash K \rangle, & $K\subset Z\backslash R$ \cr
0,& $K\not\subset Z\backslash R$ \cr} $$
correspond to HF determinant $|R\rangle $
and single parameter $\bar C_R$ is used to ensure its proper normalization.

It is pertinent to note that in the frameworks of this scheme families of
$\varkappa$-electrons functions are used to perform local parametrization
of subsets from $J_{n,p,q}(Z)$. Of course, the cases of small $\varkappa$ 
and 
$q=n-p$ are of primary interest. The equality $q=n-p$ means that $Z=N$
on the level $q$, ${\cal S}_{n,p,q}\sim {\cal F}_{n,p}$ and, consequently,
the above scheme is applied, in fact, for parametrization of $p$-electron 
states. If $n-p>p$ then it is necessary to turn to the hole representation of
wavefunctions and operators and work with co-sheaves ($(\bar p,q)$-sheaves with 
$\bar p=n-p$). Since $n-p>p$ implies $\bar p >n-\bar p$, for co-sheaves 
the equality $q=n-\bar p$ can certainly be reached. 
  
We illustrate our approach  with one  relatively simple but somewhat 
formal example. Let 
$${\psi}_Z={{{p+q-\varkappa}\choose p}}
\sum\limits_{S\subset Z}^{(\varkappa)}c_S|S\rangle \eqno(49)$$
be some trial $\varkappa$-electron function expanded over ${\varkappa}$-electron
determinants with indices from the index set $Z,|Z|=p+q$. Rewriting it in 
the form  
$${\psi}_Z=
\sum\limits_{K\subset Z}^{(q-\varkappa)}
\sum\limits_{S\subset Z\backslash K}^{(\varkappa)} 
c_S|S\rangle =$$
$$=\sum\limits_{K\subset Z}^{(q-\varkappa)}
\sum\limits_{R\subset Z\backslash K}^{(p)} 
(-1)^{|R\cap {\Delta}_{Z\backslash K}|}\bar C_R^{(K)}|Z\backslash K\backslash R
\rangle ,\eqno(50)$$
where $\bar C_R^{(K)}=
(-1)^{|R\cap {\Delta}_{Z\backslash K}|}
c_{Z\backslash K\backslash R}$,
and replacing each   $K$-component in expansion (50) 
by the corresponding simple sheaf,
we arrive at the $(p,\varkappa)$-sheaf of the form of Eq.(48) 
depending on ${{p+q}\choose {\varkappa}}$ free $\varkappa$-electron parameters. 
{\sl Sheaf thus  constructed will be called $(p,\varkappa)$-sheaf induced by 
$\varkappa$-electron wavefunction (49) and will be denoted as 
$[{\psi}_Z]_{p,\varkappa}$.} It is easy to see that arbitrary linear combination of 
induced sheaves is induced sheaf.

The next step is to derive the explicit formulas for the matrix elements of 
the $\varkappa$-electron Hamiltonian $H_{p,q,\varkappa}(Z)$ corresponding to 
the energy expression 
$$E_Z=\frac{1}{{{n-p}\choose \varkappa}}\sum\limits_{K,K'\subset Z}^{(q-\varkappa)}
\sum\limits_{Z'\subset N}^{(\varkappa)}
\langle {\psi}_{Z\backslash KZ'}|A^{-1}(n,p,\varkappa)H_{p\to \varkappa}|     
{\psi}_{Z\backslash K'Z'}\rangle .\eqno(51)$$
This complicated and somewhat tiresome procedure is described in Appendix A.

{\sl Since the inner product of $(p,\varkappa)$-sheaves (see Eq.(8)) is consistent 
with the inner product of the corresponding $p$-electron wavefunctons, it  
comes as no surprise that the eigenvalue problem for the 
$\varkappa$-electron Hamiltonian obtained from Eq.(51) should be solved in a 
certain $p$-electron metric}. 

With our choice of $\varkappa$-electron functions 
${\psi}_{Z\backslash K}$ (see Eq.(50)) the coefficients in expansion of the
corresponding $p$-electron wavefunction (30) are 
$$\sum\limits_{K\subset Z\backslash R}^{(q-\varkappa)}\bar C^{(K)}_R=
(-1)^{|R\cap {\Delta}_R|}
\sum\limits_{S\subset Z\backslash R}^{(\varkappa)}
(-1)^{|R\cap {\Delta}_S|}c_S,$$
and the normalization condition for $\varkappa$-electron coefficients $c_S$ 
takes the form
$$\sum\limits_{S,S'\subset Z}^{(\varkappa)}\left (
\sum\limits_{R\subset Z\backslash (S\cup S')}^{(p)}
(-1)^{|R\cap {\Delta}_{S\Delta S'}|}\right ) c^*_Sc_{S'}=1,$$
where the comparision $|R\cap {\Delta}_S|+|R\cap {\Delta}_{S'}|\equiv 
|R\cap {\Delta}_{S\Delta S'}| (\mbox{\rm mod}\ 2)$ was taken into account 
(see Appendix A of {\cite {Panin-3}}).

Let us introduce the metric matrix
$$\left [G_{p,q,\varkappa}(Z)\right ]_{SS'}=
\sum\limits_{R\subset Z\backslash (S\cup S')}^{(p)}
(-1)^{|R\cap {\Delta}_{S\Delta S'}|}.\eqno(52)$$

It is easy to show that the matrix (52) is nonnegative. However, it is not
necessarily strictly positive. In Appendix B it is demonstrated how to 
derive the expression for the matrix elements of 
$G_{p,q,\varkappa}(Z)$ as  a sum of binomials.  

Matrices $G_{p,q,\varkappa}(Z)$ are of purely combinatorial nature 
(see Appendix B) and to study their properties methods of modern enumerative
combinatorics should most likely be used. The explicit description of 
kernals of these  matrices is of primary interest. 
It seems to be an interesting combinatorial problem with 
unexpectedly simple and elegant solution. Indeed, 
numerical experiments led us to the following two hypothesis concerning the 
properties of these matrices.

{\bf Hypothesis 1.} 
$$\dim_{\Bbb C} \ker\left [G_{p,q,\varkappa}(Z)\right ]=\cases{
{{{p+q-1}\choose {\varkappa-1}}}  &if $q-\varkappa\equiv 1 (\mbox{\rm mod}\ 2)$,\cr
\ \ \ \ 0                         &if $q-\varkappa\equiv 0 (\mbox{\rm mod}\ 2)$.\cr}
$$

{\bf Hypothesis 2.} If $q\equiv 0 (\mbox {\rm mod}\ 2)$ then
$$ \ker\left [G_{p,q,1}(Z)\right ]=\Bbb C \sum\limits_{s\in Z}
(-1)^s|\{s\}\rangle. $$

If the Hypothesis 1 is true, then for $q-\varkappa\equiv 0 (\mbox{\rm mod}\ 2)$
matrices $G_{p,q,\varkappa}(Z)$ are strictly positive and, consequently, 
in this case the minimization of energy (51) reduces to the 
solution of the standard generalized eigenvalue problem in the 
$\varkappa$-electron space equipped with $p$-electron metric:
$$H_{p,q,\varkappa}(Z){\psi}_Z=E_ZG_{p,q,\varkappa}(Z){\psi}_Z,\eqno(53)$$   
Explicit expressions for $\varkappa$-electron  Hamiltonian $ H_{p,q,\varkappa}(Z)$ 
both in the particle and hole representations are given in Appendix A.   

When trial function (49) is chosen to be one-electron one (just a molecular
spin orbital) then it is necessary at first to lift its $K$-components to a 
higher level $\varkappa$. In contrast to the situation described in 
Proposition 7, in the case under discussion the index set $Z$ should stay
unchanged. One of the ways to do it consists in construction of the functions
$${\psi}_{Z\backslash K'}=\sum\limits_{R\subset Z\backslash K'}^{(p)}
\left ( \sum\limits_{K\atop {(K'\subset K\subset Z)}}^{(q-1)}
(-1)^{|R\cap {\Delta}_{Z\backslash K}|}\bar C_R^{(K)}\right )
|Z\backslash K'\backslash R \rangle, $$
where  $|K'|=q-\varkappa$.
 With such choice of $K'$-components at the level $\varkappa$ the 
 normalization condition for one-electron coefficients takes the form
$$\sum\limits_{i,j\in Z}c_i^*c_j\sum\limits_{{S\subset Z\backslash \{i\}}
\atop {S'\subset Z\backslash \{j\}}}^{(\varkappa -1)}
\left [ G_{p,q-2,\varkappa-1}(Z\backslash \{i,j\})\right ]_{SS'}=1,$$
and, consequently, at the one-electron level we are faced with the 
generalized eigenvalue problem with respect to some contracted metric.
We leave this aspect of theory to future study.

It is clear that for fixed $Z$ the number of linearly independent induced
$(p,\varkappa)$-sheaves is equal to ${{p+q}\choose {\varkappa}}-
dim_{\Bbb C} ker[G_{p,q,\varkappa}(Z)]$. One can hardly hope that linear
combinations of induced $(p,\varkappa)$-sheaves may be used directly
in molecular calculations. First, the total number of free parameters is too 
small. Second, the scheme of construction of induced sheaves does not
have any robust physical idea behind it. The situation may be partially
improved by turning to $(p,\varkappa)$-sheaf
$$\lambda s_{n,p,\varkappa}(R)+[{\psi }_Z]_{p,\varkappa},\eqno(54)$$
where
$$  s_{n,p,\varkappa}(R) =\left \{(-1)^{|R\cap {\Delta}_{Z'}|}
|Z'\backslash R \rangle \right \}_{R\subset Z'\subset N}\eqno(55)$$
is the determinant sheaf corresponding to HF $p$-electron wavefunction $|R\rangle $,
and $\lambda $ is the additional parameter to be optimized. It is easy to
demonstrate that the problem of determination of optimal parameters 
$\lambda $ and $c_S, S\subset Z$ in expression (54) may be reduced to the 
eigenvalue problem involving augmented Hamiltonian and augmented metric:
$$
\left (\matrix {E_R& b^{\dagger}_{p,q,\varkappa}(Z)\cr
b_{p,q,\varkappa}(Z)& H_{p,q,\varkappa}(Z)\cr}\right ) 
\left (\matrix {\lambda \cr
                  c\cr}\right )=
E_Z\left (\matrix {1& a^{\dagger}_{p,q,\varkappa}(Z)\cr
a_{p,q,\varkappa}(Z)& G_{p,q,\varkappa}(Z)\cr}\right ) 
\left (\matrix {\lambda \cr
                  c\cr}\right ),\eqno(56)$$

where
$$a_{p,q,\varkappa}(Z)=\frac{\partial}{\partial c}
\langle s_{n,p,\varkappa}(R)|[{\psi }_Z]_{p,\varkappa}
\rangle , \eqno(57)$$
$$b_{p,q,\varkappa}(Z)=\frac{\partial}{\partial c}
\langle s_{n,p,\varkappa}(R)|A^{-1}(n,p,\varkappa)|[{\psi }_Z]_{p,\varkappa}
\rangle , \eqno(58)$$
and $E_R$ is the HF energy of the determinant state $|R\rangle $. HF sheaf in 
expression (54) may be replaced by any available sheaf with subsequent
obvious modifications of Eqs.(56)-(58).

Use of more general parametrizations in accordance to the scheme
described at the beginning of this section is more complicated 
because it requires development of a certain strategy 
for {\sl systematic} selection of reasonable number of free parameters 
$c^{(K)}_S$. 
Dependence 
of coefficients in expansion (47) on $K$ results in appearance of 
normalization conditions of the following general type
$$\sum\limits_{S,S'\subset Z}^{(\varkappa)}\left [
\sum\limits_{R\subset Z\backslash (S\cup S')}^{(p)}
(-1)^{|R\cap {\Delta}_{S\Delta S'}|}
\left (c^{(Z\backslash R\backslash S)}_S\right )^*
c^{(Z\backslash R\backslash S')}_{S'} \right ] =1.$$
If no additional restrictions on coefficients
$c^{(K)}_S$ are imposed then  the corresponding metric matrix is of the form
$$\left [G^{(KK')}_{p,q,\varkappa}(Z)\right ]_{SS'}=
\sum\limits_{R\subset Z\backslash (S\cup S')}^{(p)}
(-1)^{|R\cap {\Delta}_{S\Delta S'}|}
\zeta_{Z\backslash R\backslash S,K}
\zeta_{Z\backslash R\backslash S',K'},\eqno(59)$$
where $\zeta$ is combinatorial zeta-function (see Appendix A).
In fact, the sum on the right-hand side of Eq.(59) may be equal either to $0$,
or to $\pm 1$.

As an example one can consider parametrization that is a direct
generalization of the HF case: 
$${\psi}_{Z\backslash K}=\cases { \sum\limits_{S\subset Z\backslash K}
^{(\varkappa)}c^{(K)}_S|S\rangle & $K\subset Z\backslash R$, \cr
0,& $K\not\subset Z\backslash R$ \cr} $$
The number of free parameters here is ${{p+\varkappa}\choose {\varkappa}}
{q\choose \varkappa}$.

Parametrizations  with the total number of free 
parameters   equal to the dimension of the set $Ens({\cal S}_{n,p,2}(Z))$
are of primary interest. 
On this route we may hope to approach the exact solution of $p$-electron problem by 
$\varkappa$-electron means that is the primary goal of the representability 
theory.  

{\sl It seems pertinent to emphasize once again that equations of the type of Eq.(53) 
and Eq.(56) are $\varkappa$-electron 
equations and in contrast to the standard configuration interaction  method
no expansions over $p$-electron determinants (states) appear in the 
frameworks of such an approach.}
\bigbreak
\hrule
\bigbreak
  
{\bf Orbital Representation}
\bigbreak
\bigbreak

Molecular calculations are normally performed  in orthonormal 
one-electron bases of molecular orbitals (MO). 
Turning from MSO to MO basis
leads to somewhat cumbersome formulas but simplifies concrete
calculations. Following Handy {\cite{Handy}} , we replace  spin-orbital index 
set $N$ by a pair of orbital index sets $(M,M)$ of $\alpha$ and 
$\beta$ spins assuming that the spatial parts of molecular 
spin-orbitals with the same orbital index are identical. This corresponds 
to the so-called 'restricted' theories whereas general MSO basis embraces
in addition 'unrestricted' theories.    
With such an approach each $(p+q)$-element subset $Z\subset N$
should be replaced by the pair $(Z_{\alpha},Z_{\beta})\subset M\times M$ with
$|Z_{\alpha}|+|Z_{\beta}|=p+q$. 

If the total number p of electrons is fixed and no other restrictions are 
imposed, then there is no essential difference between MSO and MO 
representations. For example, the expression for the metric matrix in terms
of molecular orbitals may be easily  derived from Eq.(52) and Eq.(C.1):
$$\bigl [G_{p,q,\varkappa}(Z_{\alpha},Z_{\beta})\bigr ]_
{(S_{\alpha},S_{\beta})(S_{\alpha}',S_{\beta}')}$$
$$=\sum\limits_{{p_{\alpha},p_{\beta}}\atop{(p_{\alpha}+p_{\beta}=p)}}
(-1)^{p_{\alpha}({\varkappa}_{\beta}+{\varkappa}_{\beta}')}
\bigl [G_{p_{\alpha},q_{\alpha},{\varkappa}_{\alpha},{\varkappa}_{\alpha}'}
(Z_{\alpha})\bigr ]_{S_{\alpha}S_{\alpha}'} 
\bigl [G_{p_{\beta},q_{\beta},{\varkappa}_{\beta},{\varkappa}_{\beta}'}(Z_{\beta})
\bigr ]_
{S_{\beta}S_{\beta}'},\eqno(60)$$
where 
$${\varkappa}_{\alpha}=|S_{\alpha}|, {\varkappa}_{\beta}=|S_{\beta}|,
{\varkappa}_{\alpha}+{\varkappa}_{\beta}=\varkappa,$$
$${\varkappa}_{\alpha}'=|S_{\alpha}'|, {\varkappa}_{\beta}'=|S_{\beta}'|,
{\varkappa}_{\alpha}'+{\varkappa}_{\beta}'=\varkappa,$$
$$q_{\alpha}=|Z_{\alpha}|-p_{\alpha}, q_{\beta}=|Z_{\beta}|-p_{\beta},$$
$$|Z_{\alpha}|+|Z_{\beta}|=p+q.$$

If, in addition, the projection $M_S$ of the total spin
is fixed, then the numbers $p_{\alpha}$ of ${\alpha}$ and $p_{\beta}$ of 
${\beta}$ electrons are also fixed for each split basis determinant 
$|R_{\alpha},R_{\beta}\rangle $. 
Moreover, the $q$ value for a
given pair $(Z_{\alpha},Z_{\beta})\subset M\times M$ can be uniquely
presented as a sum of its $\alpha$ and $\beta$ components: 
$$q=|Z_{\alpha}\backslash R_{\alpha}|+|Z_{\beta}\backslash R_{\beta}|=
q_{\alpha}+q_{\beta}.$$

{\bf Definition.} Pair $(Z_{\alpha},Z_{\beta})$ is called a pair of 
index $(q_{\alpha},q_{\beta})$ if $|Z_{\alpha}|=p_{\alpha}+q_{\alpha}$ and 
$|Z_{\beta}|=p_{\beta}+q_{\beta}$.  

Thus, in the orbital representation for the fixed value of the total spin projection 
there are $q+1$ different types of pairs 
$(Z_{\alpha},Z_{\beta})$ and arbitrary $(p,q)$-sheaf 
$\left \{{\psi}_{(Z_{\alpha},Z_{\beta})}\right \}_{(Z_{\alpha},Z_{\beta})\subset M\times M}$
involves, in general, $q$-germs with labels $(Z_{\alpha},Z_{\beta})$ of 
different index.

In MO basis the expressions for $(p,q)$-sheaves become rather 
complicated (see Appendix C). Here we discuss only MO representation of 
the metric matrix and the Hamiltonian involved in Eqs.(53) and (56).

Simple combinatorial arguments lead to the conclusion that 
$(K_{\alpha},K_{\beta})$
component of arbitrary $\varkappa$-electron wavefunction 
${\psi_{(Z_{\alpha},Z_{\beta})}}$    
corresponding to some given  $M_S$ value is expanded over determinants 
$|S_{\alpha},S_{\beta}\rangle $ with {\it fixed} $|S_{\alpha}|=
{\varkappa}_{\alpha}=q_{\alpha}-|K_{\alpha}|$ 
and $|S_{\beta}|={\varkappa}_{\beta}=q_{\beta}-|K_{\beta}|$. As a result, 
trial $\varkappa$-electron   
wavefunction (see Eq.(49)) should be taken in the form
$${\psi}_{(Z_{\alpha},Z_{\beta})}^{M_S}
=\sum\limits_{{\varkappa}_{\alpha}=max(0,\varkappa -q_{\beta})}
^{min(\varkappa, q_{\alpha})}
$$
$$\times \sum\limits_{S_{\alpha}\subset Z_{\alpha}}
^{({\varkappa}_{\alpha})}\sum\limits_{S_{\beta}\subset Z_{\beta}}
^{({\varkappa}_{\beta})}{{p_{\alpha}+q_{\alpha}-{\varkappa}_{\alpha}}\choose p_{\alpha}}
{{p_{\beta}+q_{\beta}-{\varkappa}}_{\beta}\choose p_{\beta}}
c_{(S_{\alpha},S_{\beta})}|S_{\alpha},S_{\beta}\rangle ,\eqno(61)$$
where ${\varkappa }_{\beta}=\varkappa-{\varkappa }_{\alpha}$.

The metric matrix corresponding to such choice of trial $\varkappa$-electron 
wavefunction is 
$$\bigl [G_{p,q,\varkappa}^{M_S}(Z_{\alpha},Z_{\beta})\bigr ]_
{(S_{\alpha},S_{\beta})(S_{\alpha}',S_{\beta}')}$$
$$=(-1)^{p_{\alpha}({\varkappa}_{\beta}+{\varkappa}_{\beta}')}
\bigl [G_{p_{\alpha},q_{\alpha},{\varkappa}_{\alpha},{\varkappa}_{\alpha}'}
(Z_{\alpha})\bigr ]_{S_{\alpha}S_{\alpha}'} 
\bigl [G_{p_{\beta},q_{\beta},{\varkappa}_{\beta},{\varkappa}_{\beta}'}(Z_{\beta})
\bigr ]_
{S_{\beta}S_{\beta}'},\eqno(62)$$

where $p_{\alpha}=\frac{1}{2}(p+2M_S)$ and $p_{\beta}=\frac{1}{2}(p-2M_S)$. 

In contrast to general metric matrix (60) its $M_S$ components (62) 
turn out to be always degenerate.

$M_S$ component of $\varkappa$-electron Hamiltonian $H_{p,q,\varkappa}(Z)$ 
in the orbital representation is easily obtained from Eq.(A.12) with the
aid of the relation (C.1):
$$(-1)^{p_{\alpha}({\varkappa}_{\beta} +{\varkappa }_{\beta} ')}
\bigl [H_{p,q,\varkappa}^{M_S} \bigr ]_{(S_{\alpha},S_{\beta})
(S_{\alpha}',S_{\beta}')}=  $$
$$= \left \{\sum\limits_{{j\in Z_{\alpha}\backslash S_{\alpha}}\atop 
{i\in Z_{\alpha}\backslash S_{\alpha}'}}
(-1)^{|\{j\}\cap {\Delta}_{S_{\alpha}}|+|\{i\}\cap {\Delta}_{S_{\alpha}'}|}
(i|h|j)\bigl [G_{p_{\alpha} -1,q_{\alpha} +1,{\varkappa}_{\alpha}+1,
{\varkappa}_{\alpha}'+1}(Z_{\alpha})\bigr ]_{S_{\alpha}\cup \{j\}
S_{\alpha}'\cup \{i\}}\right \}$$
$$\times \bigl [G_{p_{\beta},q_{\beta},{\varkappa}_{\beta},
{\varkappa}_{\beta}'}(Z_{\beta})\bigr ]_{S_{\beta}S_{\beta}'}$$
$$+ \left \{\sum\limits_{{j\in Z_{\beta}\backslash S_{\beta}}\atop 
{i\in Z_{\beta}\backslash S_{\beta}'}}
(-1)^{|\{j\}\cap {\Delta}_{S_{\beta}}|+|\{i\}\cap {\Delta}_{S_{\beta}'}|}
(i|h|j)\bigl [G_{p_{\beta} -1,q_{\beta} +1,{\varkappa}_{\beta}+1,
{\varkappa}_{\beta}'+1}(Z_{\beta})\bigr ]_{S_{\beta}\cup \{j\}
S_{\beta}'\cup \{i\}} \right \} $$
$$\times \bigl [G_{p_{\alpha},q_{\alpha},{\varkappa}_{\alpha},
{\varkappa}_{\alpha}'}(Z_{\alpha})\bigr ]_{S_{\alpha}S_{\alpha}'}
$$
$$
+\frac{1}{2}\left \{\sum\limits_{{{k,l\in Z_{\alpha}\backslash S_{\alpha}}
\atop {i,j\in Z_{\alpha}\backslash S_{\alpha}'}}\atop {(k\ne l, i\ne j)}}
(-1)^{|\{j\}\cap {\Delta}_{\{i\}}|+|\{l\}\cap {\Delta}_{\{k\}}|+
|\{k,l\}\cap {\Delta}_{S_{\alpha}}|+|\{i,j\}\cap {\Delta}_{S_{\alpha}'}|}
(ik|jl)\right .$$
$$\left .\times \bigl [G_{p_{\alpha} -2,q_{\alpha} +2,{\varkappa}_{\alpha}+2,
{\varkappa}_{\alpha}'+2}(Z_{\alpha})\bigr ]_{S_{\alpha}\cup \{k,l\}
S_{\alpha}'\cup \{i,j\}}
\vphantom {\sum\limits_{{{k,l\in Z_{\alpha}\backslash S_{\alpha}}\atop 
{i,j\in Z_{\alpha}\backslash S_{\alpha}'}}\atop {(k\ne l, i\ne j)}}}
\right \}
\bigl [G_{p_{\beta},q_{\beta},{\varkappa}_{\beta},
{\varkappa}_{\beta}'}(Z_{\beta})\bigr ]_{S_{\beta}S_{\beta}'}
$$
$$+\frac{1}{2}\sum\limits_{
{{{{k\in Z_{\alpha}\backslash S_{\alpha}}
\atop {l\in Z_{\beta}\backslash S_{\beta}}}
\atop {i\in Z_{\alpha}\backslash S_{\alpha}'}}
\atop {j\in Z_{\beta}\backslash S_{\beta}'}}}
(-1)^{|\{k\}\cap {\Delta}_{S_{\alpha}}|+
|\{l\}\cap {\Delta}_{S_{\beta}}|+
|\{i\}\cap {\Delta}_{S_{\alpha}'}|+
|\{j\}\cap {\Delta}_{S_{\beta}'}|}
(ik|jl)$$
$$\times 
\bigl [G_{p_{\alpha} -1,q_{\alpha} +1,{\varkappa}_{\alpha}+1,
{\varkappa}_{\alpha}'+1}(Z_{\alpha})\bigr ]_{S_{\alpha}\cup \{k\}
S_{\alpha}'\cup \{i\}}
\bigl [G_{p_{\beta} -1,q_{\beta} +1,{\varkappa}_{\beta}+1,
{\varkappa}_{\beta}'+1}(Z_{\beta})\bigr ]_{S_{\beta}\cup \{l\}
S_{\beta}'\cup \{j\}}$$
$$+\frac{1}{2}\sum\limits_{
{{{{k\in Z_{\beta}\backslash S_{\beta}}
\atop {l\in Z_{\alpha}\backslash S_{\alpha}}}
\atop {i\in Z_{\beta}\backslash S_{\beta}'}}
\atop {j\in Z_{\alpha}\backslash S_{\alpha}'}}}
(-1)^{|\{k\}\cap {\Delta}_{S_{\beta}}|+
|\{l\}\cap {\Delta}_{S_{\alpha}}|+
|\{i\}\cap {\Delta}_{S_{\beta}'}|+
|\{j\}\cap {\Delta}_{S_{\alpha}'}|}
(ik|jl)$$
$$\times 
\bigl [G_{p_{\alpha} -1,q_{\alpha} +1,{\varkappa}_{\alpha}+1,
{\varkappa}_{\alpha}'+1}(Z_{\alpha})\bigr ]_{S_{\alpha}\cup \{l\}
S_{\alpha}'\cup \{j\}}
\bigl [G_{p_{\beta} -1,q_{\beta} +1,{\varkappa}_{\beta}+1,
{\varkappa}_{\beta}'+1}(Z_{\beta})\bigr ]_{S_{\beta}\cup \{k\}
S_{\beta}'\cup \{i\}}$$
$$
+\frac{1}{2}\left \{\sum\limits_{{{k,l\in Z_{\beta}\backslash S_{\beta}}\atop 
{i,j\in Z_{\beta}\backslash S_{\beta}'}}\atop {(k\ne l, i\ne j)}}
(-1)^{|\{j\}\cap {\Delta}_{\{i\}}|+|\{l\}\cap {\Delta}_{\{k\}}|+
|\{k,l\}\cap {\Delta}_{S_{\beta}}|+|\{i,j\}\cap {\Delta}_{S_{\beta}'}|}
(ik|jl)\right .$$
$$\left .\times \bigl [G_{p_{\beta} -2,q_{\beta} +2,{\varkappa}_{\beta}+2,
{\varkappa}_{\beta}'+2}(Z_{\beta})\bigr ]_{S_{\beta}\cup \{k,l\}
S_{\beta}'\cup \{i,j\}}
\vphantom {\sum\limits_{{{k,l\in Z_{\alpha}\backslash S_{\alpha}}\atop 
{i,j\in Z_{\alpha}\backslash S_{\alpha}'}}\atop {(k\ne l, i\ne j)}}}
\right \}
\bigl [G_{p_{\alpha},q_{\alpha},{\varkappa}_{\alpha},
{\varkappa}_{\alpha}'}(Z_{\alpha})\bigr ]_{S_{\alpha}S_{\alpha}'}.
\eqno(63)$$

 The analogous expression for this Hamiltonian in the hole representation
 may easily be obtained from Eq.(A.16).
\bigbreak
\hrule
\bigbreak
\newpage
{\bf Numerical Examples}
\bigbreak

To test numerous complicated formulas obtained in the preceding sections 
and to illustrate our approach, we performed several ground state calculations
for small atomic and molecular systems using Eqs.(56)-(58). All HF calculations
to generate MO basis and list of transformed molecular integrals were carried out
with the aid of the GAMESS program {\cite {GAMESS}}.

Modification of the Davidson diagonalization routine {\cite {Davidson}}
to handle eigenvalue problem in arbitrary non-degenerate metric (generalized
eigenvalue problem) is not complicated and straightforward. We had certain
doubts about possibility to use this routine for the case of degenerate 
metrics. To our surprise, after incorporation in Gram-Schmidt orthogonalization
routine outflow for kernal vectors of the current metric we obtained 
stable working diagonalization procedure.  It comes as no surprise that 
its convergence 
properties are worse than that of the original Davidson method for the 
standard Euclidean metric but it normally manifests itself only  in the number
of iterations that are necessary to reach the solution.
  
In Table I the results of calculations of few simple systems in STO-3G
basis are presented. For $B,H_2O,NH_2,$ and $NH_3$ the  particle representation 
was used $(p\ge 2m-p)$ whereas for $Be,$ and $LiH$ transformation to the hole
representation was required $(p<2m-p)$. In all calculations $q$ value was 
taken equal to $2m-p$ and, consequently, only the case $(Z_{\alpha},Z_{\beta})=
(M,M)$ was under consideration. 

In the basis of two-electron induced sheaves $(\varkappa =2)$
only small percent of correlation energy is accounted (see Table II).
This percent normally rises when $\varkappa$ increases, and FCI limit corresponds to
$\varkappa =q$ in full accordance with general theory. It is interesing to note that
for boron atom the energy corresponding to $\varkappa =4$ is greater than that for
$\varkappa =3$. It is most probably  connected with high degeneracy  of the metric matrix
for $\varkappa =4$ (its kernal dimension is equal to 35 for $\varkappa =3$
and 75 for $\varkappa =4$). As a result, in small basis used, the actual 
number of free  parameters in 
energy turns out to be essentially less for $\varkappa =4$ in compare with     
the case $\varkappa =3$. 

We did not try to calculate more extensive systems or use better bases because
the  parametric induced sheaves even if they are 
centered at the HF origin (see Eq.(54)) will hardly become a working tool
of the computational quantum chemistry. In spite of  mathematical
beauty of metric matrices given by Eq.(B.6) and (B.7) the induced sheaves 
have no reasonable physical interpretation  and for small $\varkappa $
include too small number of parameters to be adjusted. {\sl We used them 
just to illustrate the principal possibility to solve $p$-electron problems
by diagonalization of $\varkappa $-electron Hamiltonians in  specially selected 
$p$-electron metrics.}      
\bigbreak
\hrule
\bigbreak
{\bf Conclusion}
\bigbreak

We have presented a new way to approach many electron problems using 
$\varkappa $-electron means $(\varkappa =2,3,\ldots )$. With the aid of
special parametrizations it is possible to reduce $p$-electron optimization 
problem to generalized eigenvalue problem for Hamiltonians in spaces 
corresponding to numbers of electrons less than $p$. These Hamiltonians and 
related metric matrices are uniquely determined by the parametrization
selected. Systematic study of different parametrizations and their usefullness
for concrete calculations is still an opened field for investigation. We have
suggested two relatively simple parametrization schemes and studied in 
detail one of these schemes based on the notion of the induced sheaves. In spite
of the fact that the induced sheaves  are introduced is a very formal way
and do not have any  physical idea behind their definition, they may be 
used to analyze $\varkappa $-electron contributions $(\varkappa =2,3,\ldots, q )$ 
to  correlation energy.  
     
There exists an opinion shared by a number of notable scientists that the romantic
period of quantum chemistry came to its end with the appearance of universal 
computer programs such as GAUSSIAN, MOLCAS, GAMESS, etc. As we hope,
the present paper demonstrates that it is fortunately not yet true.
 
\bigbreak
\hrule
\bigbreak
{\bf Appendix A}
\bigbreak
Here we describe a general scheme of evaluation of matrix elements 
of products of  the creation-annihilation operators involved in the 
energy expressions of the type of Eq.(51).

Let ${\psi}_Z$ be $\varkappa$-electron trial wave function expanded over 
determinants with indices from MSO index set $Z, |Z|=p+q$. Rewriting this
function as a sum of its $K$-components (see Eq.(50)) 
$${\psi}_Z=\sum\limits_{K\subset Z}^{(q-\varkappa)}{\psi}_{Z\backslash K}$$
and replacing each component in this sum by the corresponding simple 
$(p,\varkappa)$-sheaf $\{{\psi}_{Z\backslash KZ'}\}_{Z\subset N}$ with germs
$$ {\psi}_{Z\backslash KZ'}=\sum\limits_{R\subset Z'\cap (Z\backslash K)}
^{(p)}(-1)^{|R\cap {\Delta}_{Z'}|}\bar C^{(K)}_R|Z'\backslash R\rangle, $$
where $\bar C^{(K)}_R=
(-1)^{|R\cap {\Delta}_{Z\backslash K}|}c_{Z\backslash K\backslash R}$,
we come to the problem of evaluation of matrix elements
$$\langle {\psi}_Z|\mathfrak{a}|{\psi}_Z \rangle _G=
\frac{1}{{{n-p}\choose \varkappa}}$$
$$\times \sum\limits_{K,K'\subset Z}^{(q-\varkappa)}\sum\limits_{Z'\subset N}
^{(p+\varkappa)}
\sum\limits_{{R\subset Z'\cap (Z\backslash K)}\atop 
{R'\subset Z'\cap (Z\backslash K')}}^{(p)}
(-1)^{|R\cap {\Delta}_{Z'}|+|R'\cap {\Delta}_{Z'}|}\left (\bar C^{(K)}_R\right )^*
\bar C^{(K')}_{R'}\langle Z'\backslash R|\mathfrak{a}|Z'\backslash R'\rangle ,
\eqno(A.1) $$
where  $\mathfrak{a}$ stands for some relevant product of the creation-annihilation 
operators and subscript $G$ means that the inner product is taken in the metric
determined by  the matrix $G_{p,q,\varkappa}(Z)$ (see Eq.(52)). 

Changing the order of summations on the right-hand side of Eq.(A.1) and using
technque of manipulations with phase prefactors developed in {\cite{Panin-3}}, we can
recast Eq.(A.1) as
$$\langle {\psi}_Z|\mathfrak{a}|{\psi}_Z \rangle _G=
\frac{1}{{{n-p}\choose \varkappa}}
\sum\limits_{S,S'\subset Z}^{(\varkappa)}c^*_Sc_{S'} \left \{
\sum\limits_{{R\subset Z\backslash S}\atop {R'\subset Z\backslash S'}}^{(p)}
(-1)^{|R\cap {\Delta}_{S}|+|R'\cap {\Delta}_{S'}|} \right .$$
$$\left .\times \sum\limits_{Z'\supset R\cup R'}^{(p+\varkappa)}
(-1)^{|R\cap {\Delta}_{Z'\backslash R}|+|R'\cap {\Delta}_{Z'\backslash R'}|}
\langle Z'\backslash R|\mathfrak{a}|Z'\backslash R'\rangle \right \}
.\eqno(A.2)$$

There are five cases to be analyzed (see Eqs.(40a)-(40e)).

Case 1: $\mathfrak{a}=a_i^{\dagger}a_i.$

In this case the inner sum on the right-hand size of Eq.(A.2) is equal to
$$(1-{\zeta}_{\{i\},R})\sum\limits_{Z'\supset R}^{(p+\varkappa)}{\zeta}_{\{i\},Z'}=
(1-{\zeta}_{\{i\},R}){{{n-p-1}\choose {\varkappa-1}}},$$
where 
$${\zeta}_{I,R}=\cases {1 &if  $I\subset R$,\cr
                       0 &if $I\not\subset R$ \cr}$$
is the well-known in combinatorics zeta-function of partially ordered by 
inclusion set of all subsets of N {\cite{Stanley}}. By an abuse of notation, we agree 
to use ${\zeta}_{i,R}$ instead of ${\zeta}_{\{i\},R}$ for  
one-element subsets of N. In this case this function is identical to the 
characteristic function of subset $R$ accepting the 
value 1 if $i\in R$ and 0 if $i\notin R$.

If $i\in S\cup S'$ or $i\in N\backslash Z$ then the sum over 
$R\subset Z\backslash (S\cup S')$
is just proportional to the matrix element $[G_{p,q,\varkappa}]_{SS'}$.
If $i\in Z\backslash (S\cup S')$ then this sum is proportional to the 
matrix element $[G_{p,q-1,\varkappa}(Z\backslash \{i\}]_{SS'}$.
The final expression for the desired matrix element is obtained with the 
aid of the recurrence relation (B.9): 
$$\langle {\psi}_Z|a^{\dagger}_ia_i|{\psi}_Z \rangle _G=
\frac {{\varkappa\choose 1}}{{{n-p}\choose 1}}\left \{
\vphantom {\sum\limits_{S,S'\in Z}^{(\varkappa)}}
\langle {\psi}_Z|G_{p,q,\varkappa}(Z)|{\psi}_Z \rangle \right .$$ 
$$ \left .- \sum\limits_{S,S'\in Z}^{(\varkappa)} c^*_Sc_{S'} 
(-1)^{|\{i\}\cap {\Delta}_{S\Delta S'}|}
{\zeta}_{i,Z\backslash (S\cup S')}
[G_{p-1,q+1,\varkappa +1}(Z)]_
{S\cup \{i\}S'\cup \{i\}} \right \}\eqno(A.3)$$

Case 2: $\mathfrak{a}=a_i^{\dagger}a_j, i\ne j.$

The sum over $Z'$ on the right-hand side of Eq.(A.2) is non-zero only for 
$R=R_1\cup \{j\}$ and $R'=R_1\cup \{i\}$. Substitution of expansion (40b) 
for the operator $a_i^{\dagger}a_j$ in matrix elements 
$\langle Z'\backslash R|a_i^{\dagger}a_j|Z'\backslash R'\rangle$ 
makes it easy to get the following expression for the sum under consideration: 
$$(-1)^{|R_1\cap {\Delta}_{\{i,j\}}|+|\{i\}\cap {\Delta_{\{j\}}}|
+|\{j\}\cap {\Delta_{\{i\}}}|}{{{n-p-1}\choose {\varkappa-1}}}.$$
Since $|\{i\}\cap {\Delta_{\{j\}}}|+|\{j\}\cap {\Delta_{\{i\}}}| \equiv 1
({\rm mod}\ 2)$ for $i\ne j$, the required matrix element takes the form
$$\langle {\psi}_Z|a^{\dagger}_ia_j|{\psi}_Z \rangle _G=
-\frac {{\varkappa\choose 1}}{{{n-p}\choose 1}}
\sum\limits_{S,S'\in Z}^{(\varkappa)}c^*_Sc_{S'} $$
$$\times \sum\limits_{{R_1\cup \{j\}\subset Z\backslash S}
\atop {R_1\cup \{i\}{\subset Z\backslash S'}}}^{(p)}
(-1)^{|(R_1\cup \{j\})\cap{\Delta}_{S}|+|R_1\cup \{i\}\cap {\Delta}_{S'}|+
|R_1\cap {\Delta_{\{i,j\}}}|}.$$
Taking into account that $R_1\cup \{j\}\subset Z\backslash S$ implies
$j \notin S $, and $R_1\cup \{i\}\subset Z\backslash S'$ implies
$i \notin S' $,  we can write down the final expression:
$$\langle {\psi}_Z|a^{\dagger}_ia_j|{\psi}_Z \rangle _G=
-\frac {{\varkappa\choose 1}}{{{n-p}\choose 1}}
\sum\limits_{S,S'\in Z}^{(\varkappa)}c^*_Sc_{S'}
(-1)^{|\{j\}\cap {\Delta}_S|+|\{i\}\cap {\Delta}_{S'}|}$$
$$\times {\zeta}_{j,Z\backslash S}{\zeta}_{i,Z\backslash S'}
\left [G_{p-1,q+1,\varkappa +1}(Z)\right ]_{S\cup \{j\}S'\cup \{i\}}.
\eqno(A.4)$$

Case 3: $\mathfrak{a}=a_i^{\dagger}a_j^{\dagger}a_ja_i, i\ne j.$

Substitution of expansion (40c) in matrix element
$\langle Z'\backslash R|a_i^{\dagger}a_j^{\dagger}a_ja_i|Z'\backslash R'\rangle$
immediately shows that non-zero contributions correspond to the case 
$R=R'$ and $R\not\ni i,j$:
$$\langle {\psi}_Z|a_i^{\dagger}a_j^{\dagger}a_ja_i|{\psi}_Z \rangle _G=
\frac {{\varkappa\choose 2}}{{{n-p}\choose 2}}
\sum\limits_{S,S'\in Z}^{(\varkappa)}c^*_Sc_{S'}
\left \{\sum\limits _{{R\subset Z\backslash (S\cup S')}\atop
{(R\not\ni i,j)}}
(-1)^{|R\cap {\Delta}_{S\Delta S'}|}\right \}.$$

Thorough analysis of mutual layout of indices $i,j$ and relevant subsets 
of the MSO index set with subsequent use of the recurrence relations (B.9)-(B.10) 
leads to the following somewhat cumbersome expression:
$$\langle {\psi}_Z|a_i^{\dagger}a_j^{\dagger}a_ja_i|{\psi}_Z \rangle _G=
\frac {{\varkappa \choose 2}}{{{n-p}\choose 2}} 
\left [ \vphantom {\frac {{\varkappa \choose 2}}{{{n-p}\choose 2}}} 
\langle {\psi}_Z|G_{p,q,\varkappa}(Z)|{\psi}_Z \rangle \right .$$
$$\left .
- \sum\limits_{S,S'\in Z}^{(\varkappa)}c^*_Sc_{S'} 
\left \{ \vphantom {\frac {{\varkappa \choose 2}}{{{n-p}\choose 2}}}
(-1)^{|\{i\}\cap {\Delta}_{S\Delta S'}|}
 {\zeta}_{i,Z\backslash (S\cup S')}[G_{p-1,q+1,\varkappa +1}(Z)]_
{S\cup \{i\}S'\cup \{i\}} \right . \right .$$ 
$$\left .\left . 
+(-1)^{|\{j\}\cap {\Delta}_{S\Delta S'}|}
{\zeta}_{j,Z\backslash (S\cup S')}[G_{p-1,q+1,\varkappa +1}(Z)]_
{S\cup \{j\}S'\cup \{j\}} \right . \right .$$
$$\left .\left .
-(-1)^{|\{i,j\}\cap {\Delta}_{S\Delta S'}|}
{\zeta}_{\{i,j\},Z\backslash (S\cup S')}[G_{p-2,q+2,\varkappa +2}(Z)]_
{S\cup \{i,j\}S'\cup \{i,j\}} 
\vphantom {\frac {{\varkappa \choose 2}}{{{n-p}\choose 2}}}\right \} 
\vphantom {\frac {{\varkappa \choose 2}}{{{n-p}\choose 2}}}\right ].\eqno(A.5)
$$

Case 4: $\mathfrak{a}=a_i^{\dagger}a_j^{\dagger}a_la_i, i\ne j\ne l.$

Substitution of expansion (40d) in matrix element 
$\langle Z'\backslash R|a_i^{\dagger}a_j^{\dagger}a_la_i|Z'\backslash R'\rangle$
readily leads to the conclusion that this matrix element is non-zero if 
$R=R_1\cup \{l\}$, $R'=R_1\cup \{j\}$, and $R_1\not\ni i$. The final expression
is
$$\langle {\psi}_Z|a_i^{\dagger}a_j^{\dagger}a_la_i|{\psi}_Z \rangle _G=
-\frac {{\varkappa\choose 2}}{{{n-p}\choose 2}}
\sum\limits_{S,S'\in Z}^{(\varkappa)}c^*_Sc_{S'}
{\zeta}_{l,Z\backslash S}{\zeta}_{j,Z\backslash S'}$$
$$\times \left \{ \vphantom  {\frac {{\varkappa\choose 2}}{{{n-p}\choose 2}}}
(-1)^{|\{l\}\cap {\Delta}_S|+|\{j\}\cap {\Delta}_{S'}|}
[G_{p-1,q+1,\varkappa+1}(Z)]_{S\cup \{l\}S'\cup \{j\}} \right .$$ 
$$\left .
-(-1)^{|\{i,l\}\cap {\Delta}_S|+|\{i,j\}\cap {\Delta}_{S'}|+
|\{i\}\cap {\Delta}_{\{j,l\}}|}{\zeta}_{i,Z\backslash (S\cup S')}
[G_{p-2,q+2,\varkappa+2}(Z)]_{S\cup \{i,l\}S'\cup \{i,j\}} 
\vphantom  {\frac {{\varkappa\choose 2}}{{{n-p}\choose 2}}} \right \}
\eqno(A.6)$$
   
Case 5: $\mathfrak{a}=a_i^{\dagger}a_j^{\dagger}a_la_k, i\ne j\ne k\ne l.$

Matrix element 
$\langle Z'\backslash R|a_i^{\dagger}a_j^{\dagger}a_la_k|Z'\backslash R'\rangle$
is nonzero only if $R=R_1\cup \{k,l\}$, $R'=R_1\cup \{i,j\}$.We have 
$$\langle {\psi}_Z|a_i^{\dagger}a_j^{\dagger}a_la_k|{\psi}_Z \rangle _G$$
$$=(-1)^{|\{j\}\cap {\Delta}_{\{i\}}|+|\{l\}\cap {\Delta}_{\{k\}}|}
\frac {{\varkappa\choose 2}}{{{n-p}\choose 2}}
 \sum\limits_{S,S'\in Z}^{(\varkappa)}c^*_Sc_{S'}
\left \{
\vphantom {\frac {{\varkappa\choose 2}}{{{n-p}\choose 2}}}
(-1)^{|\{k,l\}\cap {\Delta}_{S}|+|\{i,j\}\cap {\Delta}_{S'}|}\right .$$
$$\left . \times 
{\zeta}_{k,Z\backslash S}
{\zeta}_{l,Z\backslash S}
{\zeta}_{i,Z\backslash S'}
{\zeta}_{j,Z\backslash S'}
[G_{p-2,q+2,\varkappa +2}(Z)]_{S\cup \{k,l\}S'\cup \{i,j\}}
\vphantom {\frac {{\varkappa\choose 2}}{{{n-p}\choose 2}}}
\right \}
\eqno(A.7)$$
Now it is easy to derive the expression for the matrix elements of general
one- and two-electron operators:
$$\mathfrak{h}=\sum\limits_{i,j=1}^n\langle i|\mathfrak{h}|j\rangle 
a_i^{\dagger}a_j,\eqno(A.8)$$
$$\mathfrak{g}=\sum\limits_{i,j,k,l=1}^n\langle ij|kl\rangle 
a_i^{\dagger}a_j^{\dagger}a_la_k.\eqno(A.9)$$

We have
$$\langle {\psi}_Z|\mathfrak{h}|{\psi}_Z \rangle _G=
\frac {{\varkappa\choose 1}}{{{n-p}\choose 1}}\left \{
\vphantom {\sum\limits_{{j\in Z\backslash S}\atop {i\in Z\backslash S'}}}
Tr(h)\langle {\psi}_Z|G_{p,q,\varkappa} (Z)|{\psi}_Z \rangle\right .$$
$$\left .- \sum\limits_{S,S'\in Z}^{(\varkappa)}c^*_Sc_{S'}
\sum\limits_{{j\in Z\backslash S}\atop {i\in Z\backslash S'}}
(-1)^{|\{j\}\cap {\Delta}_S|+|\{i\}\cap {\Delta}_{S'}|}
\langle i|\mathfrak{h}|j\rangle [G_{p-1,q+1,\varkappa +1}(Z)]_
{S\cup \{j\}S'\cup \{i\}}
\right \},\eqno(A.10)$$
and
$$\langle {\psi}_Z|\mathfrak{g}|{\psi}_Z \rangle _G=
\frac {{\varkappa\choose 2}}{{{n-p}\choose 2}}\left \{
\vphantom {\frac {{\varkappa\choose 1}}{{{n-p}\choose 1}}}
\Bigl [\sum\limits_{i,j=1}^n
\bigl (\langle ij|ij\rangle - \langle ij|ji\rangle \bigr )\Bigr ]
\langle {\psi}_Z|G_{p,q,\varkappa }(Z)|{\psi}_Z \rangle \right .$$
$$\left . -2\sum\limits_{S,S'\in Z}^{(\varkappa)}c^*_Sc_{S'}
\sum\limits_{{k\in Z\backslash S}\atop {j\in Z\backslash S'}}
(-1)^{|\{k\}\cap {\Delta}_S|+|\{j\}\cap {\Delta}_{S'}|}\right .$$
$$\left .\times \Bigl [\sum\limits_{i=1}^n\bigl (\langle ij|ik\rangle - 
\langle ij|ki\rangle\bigr )\Bigr ]
[G_{p-1,q+1,\varkappa +1}(Z)]_{S\cup \{k\}S'\cup \{j\}}\right .$$
$$\left . +\sum\limits_{S,S'\in Z}^{(\varkappa)}c^*_Sc_{S'}
\sum\limits_{{{k,l\in Z\backslash S}\atop {i,j\in Z\backslash S'}}
\atop {(k\ne l,i\ne j)}}
(-1)^{|\{j\}\cap {\Delta}_{\{i\}}|+|\{l\}\cap {\Delta}_{\{k\}}|+
|\{k,l\}\cap {\Delta}_S|+|\{i,j\}\cap {\Delta}_{S'}|}\right .$$
$$\left . \times \langle ij|kl\rangle
[G_{p-2,q+2,\varkappa +2}(Z)]_{S\cup \{k,l\}S'\cup \{i,j\}}
\vphantom {\frac {{\varkappa\choose 1}}{{{n-p}\choose 1}}}
\right \}\eqno(A.11)$$
The Hamiltonian associated with the energy expression (51) is
$$[H_{p,q,\varkappa}(Z)]_{SS'}= 
\sum\limits_{{j\in Z\backslash S}\atop {i\in Z\backslash S'}}
(-1)^{|\{j\}\cap {\Delta}_S|+|\{i\}\cap {\Delta}_{S'}|}$$
$$\times \langle i|h|j\rangle [G_{p-1,q+1,\varkappa +1}(Z)]_
{S\cup \{j\}S'\cup \{i\}}$$
$$+\frac {1}{2}\sum\limits_{{{k,l\in Z\backslash S}
\atop {i,j\in Z\backslash S'}}
\atop {(k\ne l,i\ne j)}}
(-1)^{|\{j\}\cap {\Delta}_{\{i\}}|+|\{l\}\cap {\Delta}_{\{k\}}|+
|\{k,l\}\cap {\Delta}_S|+|\{i,j\}\cap {\Delta}_{S'}|}$$
$$\times \langle ij|kl\rangle
[G_{p-2,q+2,\varkappa +2}(Z)]_{S\cup \{k,l\}S'\cup \{i,j\}}\eqno(A.12)$$

As has already been mentioned, for $p<n-p$ it is reasonable to turn to
sheaves in hole representation (co-sheaves). 
Reduced hole Hamiltonian (see Appendix B of {\cite{Panin-1}}) after its transformation with 
the aid of $A^{-1}(n,\bar p,\varkappa)=A(n,p,\varkappa)$ may be written as
$$A^{-1}(n,\bar p,\varkappa)H_{\bar p \to \varkappa}^{\circ}=
\frac {{p\choose 1}}{{\varkappa\choose 1}}\sum\limits_{i,j=1}^n \overline {\langle i|h|j\rangle }
a^{\dagger}_ia_j$$ 
$$+ \frac{1}{2}\frac {{p\choose 2}}{{\varkappa \choose 2}}
\sum\limits_{i,j,k,l=1}^n \overline {\langle ij|kl \rangle}
a^{\dagger}_ia^{\dagger}_ja_la_k,\eqno(A.13)$$
where  
$$\overline {\langle i|h|j\rangle }=(-1)^{i+j}\langle j|h|i\rangle ,\eqno(A.14) $$
$$\overline {\langle ij|kl \rangle}=(-1)^{i+j+k+l}\langle kl|ij \rangle ,\eqno(A.15)$$
and  $\bar p=n-p$.
 Using Eqs.(A.10)-(A.11) it is easy to derive the 
expression for the Hamiltonian corresponding to the energy (51) in the hole 
representation:
$$[H_{p,q,\varkappa}^{\circ}(Z)]_{SS'}=
\frac{1}{2}\bigl [Tr(h+F_N)\bigr] \bigl [G_{\bar p,q,\varkappa}(Z)\bigr ]_{SS'}$$  
$$-\sum\limits_{{j\in Z\backslash S}\atop {i\in Z\backslash S'}}
(-1)^{|\{j\}\cap {\Delta}_S|+|\{i\}\cap {\Delta}_{S'}|}
\overline {\langle i|F_N|j\rangle} [G_{\bar p-1,q+1,\varkappa +1}(Z)]_
{S\cup \{j\}S'\cup \{i\}}$$
$$+\frac {1}{2}\sum\limits_{{{k,l\in Z\backslash S}\atop {i,j\in Z\backslash S'}}
\atop {(k\ne l,i\ne j)}}
(-1)^{|\{j\}\cap {\Delta}_{\{i\}}|+|\{l\}\cap {\Delta}_{\{k\}}|+
|\{k,l\}\cap {\Delta}_S|+|\{i,j\}\cap {\Delta}_{S'}|}$$
$$\times \overline {\langle ij|kl\rangle}
[G_{\bar p-2,q+2,\varkappa +2}(Z)]_{S\cup \{k,l\}S'\cup \{i,j\}},\eqno(A.16)$$
where $|Z|=\bar p+q$ and $F_N$ is the Fock operator defined by Eq.(43).
 
\bigbreak
\hrule
\bigbreak
{\bf Appendix B }
\bigbreak

With each subset $Z\subset N, |Z|=p+q$ let us associate the linear mapping 
$$g_{p,q}(Z):\bigoplus_{\varkappa =1}^q{\cal F}_{n,\varkappa}(Z) 
\to \bigoplus_{\varkappa =1}^q {\cal F}_{n,\varkappa}(Z),
\eqno(B.1)$$
defined by its matrix representation 
$$\left [G_{p,q,\varkappa,\varkappa'}(Z)\right ]_{SS'}=
\sum\limits_{R\subset Z\backslash (S\cup S')}^{(p)}
(-1)^{|R\cap {\Delta}_{S\Delta S'}|}.\eqno(B.2)$$
with respect to the determinant basis set. Here $|S|=\varkappa$, and $|S'|=
\varkappa'$. Diagonal (with respect to $\varkappa$) blocks of this matrix
are identical to the $p$-electron metric matrices given by Eq.(52).
We agree to omit repeating index $\varkappa$ in writing diagonal (with respect
to $\varkappa$) matrix elements.
Off-diagonal ($\varkappa \ne \varkappa'$) blocks appear in going from
molecular spin-orbitals to molecular orbitals.
    
Let $S$, and $S'$ be $\varkappa$- and $\varkappa'$-element subsets 
of the index set $Z\subset N,|Z|=p+q$, respectively. Let us suppose that
$$S\Delta S'=s_1<s_2<\ldots s_l,$$
$$Z\backslash (S\cap S')=z_1<z_2<\ldots <z_k ,$$     
where $l=|S\Delta S'|$, and $k=|Z\backslash (S\cap S')|=p+q-|S\cap S'|$. 

The index set $Z'=Z\backslash (S\cap S')$ may be presented as 
$$Z'=\bigcup\limits_{i=0}^l \{s_i\}\cup \left ([s_i+1,s_{i+1}-1]\cap 
Z'\right ),\eqno(B.3)$$
where $s_0=z_1-1$, $s_{l+1}=z_k+1$, and the symbol $[s,s']$ stands for the 
interval in the set $N=\{1,2,\ldots,n\}$. It is easy to show that 
$l\equiv \varkappa-\varkappa' ({\rm mod} \ 2)$. Cases of even and odd $l$ require
separate analysis, since the representation of the set ${\Delta}_{S\Delta S'}$ 
as a union of intervals depends on the parity of l (see Eq.(A.5) from 
{\cite{Panin-3}}):

$${\Delta}_{S\Delta S'}=\cases{ 
\bigcup\limits_{i=1}^{\frac{l}{2}}[s_{2i-1}+1,s_{2i}],   &if $l$ is even,\cr
\bigcup\limits_{i=0}^{[\frac{l}{2}]}[s_{2i}+1,s_{2i+1}], &if $l$ is odd.\cr}
\eqno(B.4)$$

Using decompositions (B.3)-(B.4), it is possible to express the sum on the 
right-hand side of Eq.(B.2) as a sum of binomials. We have
$$\left [G_{p,q,\varkappa,\varkappa'}(Z)\right ]_{SS'}=
\cases{
\sum\limits_{{r_1,r_2,\ldots,r_{l+1}\ge 0}
\atop {(\sum\limits_{i=1}^{l+1}r_i=p)}}
(-1)^{\sum\limits_{i=1}^{\frac{l}{2}}r_{2i}}\prod\limits_{i=1}^{l+1}
{{w_i}\choose {r_i}},&if $l$ is even, \cr
\sum\limits_{{r_1,r_2,\ldots,r_{l+1}\ge 0}
\atop {(\sum\limits_{i=1}^{l+1}r_i=p)}}
(-1)^{\sum\limits_{i=0}^{[\frac{l}{2}]}r_{2i+1}}\prod\limits_{i=1}^{l+1}
{{w_i}\choose {r_i}},&if $l$ is odd, \cr
}
\eqno(B.5)$$
where
$$w_i=|[s_{i-1}+1,s_i-1]\cap (Z\backslash (S\cap S'))|.$$
 By definition, $w_i\ge 0$ and $\sum\limits_{i=1}^{l+1}w_i=
|Z\backslash (S\cup S')|$.

Performing summations on the right-hand side of Eq.(A.5)  first over indices 
$r_i$ with odd $i$ and then over indices $r_i$ with even $i$ for the case of
even $l$, and in the backward succession for odd $l$, we arrive to
the following not very complicated expression   
$$\left [G_{p,q,\varkappa,\varkappa'}(Z)\right ]_{SS'}=
(-1)^{pl}\sum\limits_{r\ge 0}(-1)^r
{{\sum\limits_{i=0}^{[\frac{l}{2}]}w_{2i+1}}\choose {p-r}}
{{\sum\limits_{i=1}^{[\frac{l+1}{2}]}w_{2i}}\choose {r}}.\eqno(B.6)$$
In Eqs.(B.5)-(B.6) the symbol $[x]$ stands for the function  
extracting integral part of $x$.

If $\varkappa=\varkappa'$ and $S=S'$ then $l=0$ and  only  
$w_1=|[s_0+1,s_1-1]\cap (Z\backslash S)|=p+q-\varkappa$ 
differs from zero which means that  
$\left [G_{p,q,\varkappa}(Z)\right ]_{SS}={{p+q-\varkappa}\choose p}$
in full accordance with Eq.(B.2). If $\varkappa =q$ then for $S\ne S'$ 
there is no room for $p$-electron subsets in the set $Z\backslash (S\cup S')$ 
and in this case the only non-zero block is the diagonal one ($\varkappa'=q$) 
coinciding with the $q$-electron identity matrix.

If $Z=\{1,2,\ldots,p+q\}$ then the sums of interval lengths involved in the
right-hand side of Eq.(A.6) may be presented in a more transparent form.
Indeed, in this case $s_0=0$, $s_{l+1}=p+q+1$, and for even $l$
$$\sum\limits_{i=1}^{\frac{l}{2}}w_{2i}=
\sum\limits_{i=1}^l(-1)^{i}s_i-\frac{l}{2}-l_0,$$ 
$$\sum\limits_{i=0}^{\frac{l}{2}}w_{2i+1}=p+q+
\sum\limits_{i=1}^l(-1)^{i+1}s_i-\frac{l}{2}-l_1,$$ 
where $l_0$ and $l_1$ are the numbers of elements from $S\cap S'$ belonging to
'even' intervals $[s_{2i-1}+1,s_{2i}-1]$ and 'odd' intervals 
$[s_{2i}+1,s_{2i+1}-1]$, respectively. For odd $l$ we have
$$\sum\limits_{i=1}^{\frac{l+1}{2}}w_{2i}=p+q+1+
\sum\limits_{i=1}^l(-1)^{i}s_i-\frac{l+1}{2}-l_0,$$ 
$$\sum\limits_{i=0}^{\frac{l}{2}}w_{2i+1}=
\sum\limits_{i=1}^l(-1)^{i+1}s_i-\frac{l+1}{2}-l_1.$$
The final expression for the matrix elements under consideration
may be written as 
 
$$\left [G_{p,q,\varkappa,\varkappa'}(Z)\right ]_{SS'}$$
$$=\cases{
\sum\limits_{r\ge 0}(-1)^r
{{p+q+\sum\limits_{i=1}^l(-1)^{i+1}s_i-\frac{l}{2}-l_1}\choose {p-r}}
{{\sum\limits_{i=1}^l(-1)^{i}s_i-\frac{l}{2}-l_0}\choose {r}}, &if $l$ is even, \cr
\sum\limits_{r\ge 0}(-1)^r
{{p+q+1+\sum\limits_{i=1}^l(-1)^{i}s_i-\frac{l+1}{2}-l_0}\choose {p-r}}
{{\sum\limits_{i=1}^l(-1)^{i+1}s_i-\frac{l+1}{2}-l_1}\choose {r}}, &if $l$ is odd. \cr
}
\eqno(B.7)$$

Directly from the definition (B.2) various recurrence relations involving 
matrix elements of metric matrices may be derived. Here we confine ourselves to
the relations that are necessary for evaluation of relevant matrix elements.
Let us suppose that subsets $S,S'\subset Z$ are fixed and $K (|K|=k)$ is some 
subset of the set $Z\backslash (S\cup S')$. Then
$$\left [G_{p,q,\varkappa}(Z)\right ]_{SS'}$$
$$=\sum\limits_{k_1=0}^k\sum\limits_{K_1\subset K}^{(k_1)}
(-1)^{|K_1\cap {\Delta}_{S{\Delta}S'}|}
\left [G_{p-k_1,q-k+2k_1,\varkappa +k_1}((Z\backslash K)\cup K_1)\right ]
_{S\cup K_1S'\cup K_1}.\eqno(B.8)$$ 
In two most important particular cases $K=\{i\}$ and $K=\{i,j\}$  
general relation (B.8) reduces to
$$ 
\left [G_{p,q,\varkappa}(Z)\right ]_{SS'}=\left [G_{p,q-1,\varkappa}(Z\backslash \{i\})\right ]_{SS'}
$$
$$
+(-1)^{|\{i\}\cap {\Delta}_{S{\Delta}S'|}}
\left [G_{p-1,q+1,\varkappa+1}(Z)\right ]_{S\cup \{i\}S'\cup \{i\}},\eqno (B.9)
$$
$$
\left [G_{p,q,\varkappa}(Z)\right ]_{SS'}=\left [G_{p,q-2,\varkappa}(Z\backslash \{i,j\})\right ]_{SS'}
$$
$$
+(-1)^{|\{i\}\cap {\Delta}_{S{\Delta}S'|}}
\left [G_{p-1,q,\varkappa+1}(Z\backslash \{j\})\right ]_{S\cup \{i\}S'\cup \{i\}}
$$
$$
+(-1)^{|\{j\}\cap {\Delta}_{S{\Delta}S'|}}
\left [G_{p-1,q,\varkappa+1}(Z\backslash \{i\})\right ]_{S\cup \{j\}S'\cup \{j\}}
$$
$$
+(-1)^{|\{i,j\}\cap {\Delta}_{S{\Delta}S'|}}
\left [G_{p-2,q+2,\varkappa+2}(Z)\right ]_{S\cup \{i,j\}S'\cup \{i,j\}}.\eqno (B.10)
$$

\bigbreak
\hrule
\bigbreak
{\bf Appendix C }
\bigbreak

From Eq.(6) it follows that in expansion of $q$-germs phase prefactors of the
type $(-1)^{|R\cap {\Delta}_Z|}$ are involved. In  Handy 
orbital representation $R$ goes to $(R_{\alpha},R_{\beta})$ and $Z$ goes to $(Z_{\alpha},
Z_{\beta})$. Let us introduce the sets $\bar R_{\beta} = m+R_{\beta}$ and 
$\bar Z_{\beta} =m+Z_{\beta}$ with shifted by $m=|M|$ elements. We have
$R= R_{\alpha}\Delta \bar R_{\beta}$ and $Z= Z_{\alpha}\Delta \bar Z_{\beta}$.
Using technique of manipulation with sign prefactors developed in 
{\cite{Panin-3}}, we obtain
$$|R\cap {\Delta}_{Z}|\equiv |R_{\alpha}\cap {\Delta}_{Z_{\alpha}}|+
|R_{\alpha}\cap {\Delta}_{\bar Z_{\beta}}|+
|\bar R_{\beta}\cap {\Delta}_{Z_{\alpha}}|+
|\bar R_{\beta}\cap {\Delta}_{\bar Z_{\beta}}|\  ({\rm mod}\ 2).$$
From Eq.(A.5) of {\cite {Panin-3}} it follows that 
$|\bar R_{\beta}\cap {\Delta}_{Z_{\alpha}}|=0$ and 
$|R_{\alpha}\cap {\Delta}_{\bar Z_{\beta}}|\equiv |R_{\alpha}||Z_{\beta}|
({\rm mod} \ 2)$. As a 
result,  
$$|R\cap {\Delta}_{Z}|\equiv |R_{\alpha}\cap {\Delta}_{Z_{\alpha}}|+
|R_{\beta}\cap {\Delta}_{Z_{\beta}}|+ |R_{\alpha}||Z_{\beta}|({\rm mod}\ 2).
\eqno(C.1)$$
Certain assymetry of this relation is connected with the use of Handy split
determinant representation with $\alpha$ indices always going first. 

q-electron function
$${\psi}_{(Z_{\alpha},Z_{\beta})}=$$
$$(-1)^{p_{\alpha}|Z_{\beta}|} 
\sum\limits_{R_{\alpha}\subset Z_{\alpha}}^{(p_{\alpha})}
\sum\limits_{R_{\beta}\subset Z_{\beta}}^{(p_{\beta})}
(-1)^{|R_{\alpha}\cap {\Delta}_{Z_{\alpha}}|+|R_{\beta}\cap {\Delta}_{Z_{\beta}}|}
\bar C_{R_{\alpha},R_{\beta}}|Z_{\alpha}\backslash R_{\alpha},
Z_{\beta}\backslash R_{\beta}\rangle \eqno(C.2)$$
generates simple $(p,q)$-sheaf with germs
$${\psi}_{(Z_{\alpha},Z_{\beta})(Z'_{\alpha},Z'_{\beta})}=
$$
$$(-1)^{p_{\alpha}|Z'_{\beta}|}
\sum\limits_{R'_{\alpha}\subset Z'_{\alpha}\cap Z_{\alpha}}^{(p_{\alpha})}
\sum\limits_{R'_{\beta}\subset Z'_{\beta}\cap Z_{\beta}}^{(p_{\beta})}
(-1)^{|R'_{\alpha}\cap {\Delta}_{Z'_{\alpha}}|+|R'_{\beta}\cap {\Delta}_{Z'_{\beta}}|}
\bar C_{(R'_{\alpha},R'_{\beta})}|Z'_{\alpha}\backslash R'_{\alpha},
Z'_{\beta}\backslash R'_{\beta}\rangle \eqno(C.3)$$

From Eq.(C.2) it follows that in the orbital representation with fixed $M_S$ value
only $q$-electron functions ${\psi}_{(Z_{\alpha},Z_{\beta})}$
from the vector space ${\cal F}_{n,q}(Z_{\alpha},Z_{\beta})$  that are 
linear combinations of determinants $|S_{\alpha},S_{\beta}\rangle$ with $|S_{\alpha}|
=q_{\alpha}$ and $|S_{\beta}|=q_{\beta}$  
$(q_{\alpha}=|Z_{\alpha}|-p_{\alpha},q_{\beta}=|Z_{\beta}|-p_{\beta} )$ 
are of actual interest.
 
 In the orbital representation definitions (17) and (18) should be replaced by
$$I_p({\psi}_{(Z_{\alpha},Z_{\beta})})=\left \{(R_{\alpha},R_{\beta})\subset
(Z_{\alpha},Z_{\beta}): \bar C_{(R_{\alpha},R_{\beta})}\ne 0\right \};\eqno(C.4)$$
$$I_p^{\alpha}({\psi}_{(Z_{\alpha},Z_{\beta})})= \left \{R_{\alpha}\subset
Z_{\alpha}: \bar C_{(R_{\alpha},R_{\beta})}\ne 0 \mbox{\ for some\ } 
R_{\beta}\subset Z_{\beta} \right \};\eqno(C.5)$$
$$I_p^{\beta}({\psi}_{(Z_{\alpha},Z_{\beta})})= \left \{R_{\beta}\subset
Z_{\beta}: \bar C_{(R_{\alpha},R_{\beta})}\ne 0 \mbox{\ for some\ } 
R_{\alpha}\subset Z_{\alpha} \right \};\eqno(C.6)$$
$$L_{\sigma}({\psi}_{(Z_{\alpha},Z_{\beta})})=
\bigcup\limits_{R_{\sigma}\in I^{\sigma}_p({\psi}_{(Z_{\alpha},Z_{\beta})})}R_{\sigma}
\quad (\sigma=\alpha,\beta).\eqno(C.7)$$
and each simple $(p,q)$-sheaf                                         
$\left \{{\psi}_{(Z_{\alpha},Z_{\beta})(Z'_{\alpha},Z'_{\beta})}\right \}
_{(Z'_{\alpha},Z'_{\beta})\subset M\times M}$
is generated by any of its  
$${{2m-|L_{\alpha}({\psi}_{(Z_{\alpha},Z_{\beta})})|-
|L_{\beta}({\psi}_{(Z_{\alpha},Z_{\beta})})|}\choose 
{p+q-|L_{\alpha}({\psi}_{(Z_{\alpha},Z_{\beta})})|-
|L_{\beta}({\psi}_{(Z_{\alpha},Z_{\beta})})|}}
\eqno(C.8)$$
$q$-electron germs (see proposition 3). 
\bigbreak
{\bf ACKNOWLEDGMENTS}
\bigbreak
We gratefully acknowledge the Russian Foundation for Basic Research
(Grant 00-03-32943a) and Ministry of Education of RF
(Grant E00-5.0-62) for financial support of the present work.
\bigbreak
\hrule
\bigbreak

\newpage
\begin{table}[ht]
{\bf TABLE I}

{\bf Results of ground state test calculations: Total energies.}

\centering {
\begin{tabular}{ c c c c c c c }
  & \multicolumn{6}{ c }  {Energy, a.u.} \\
\cline{2-7}
Species & HF    &$\varkappa=2$  &$\varkappa=3$    &$\varkappa=4$      &$\varkappa=5$&$\varkappa=6$  \\
\hline

$Be(^1S)$         &    -14.351880     &   -14.366729 &-14.373701 &-14.403655 &   -         & -\\
$B(^2P) $         &    -24.148989     &   -24.152540 &-24.169145 &-24.161980 & -24.189265   &-\\
$LiH(^1\Sigma ^+)$&     -7.862024     &    -7.864463 & -7.868083 & -7.882402 &   -          &- \\
$NH_2(^2B_2)$     &    -54.834567     &   -54.835048 &-54.839962 &-54.850959 &  -54.882707  &- \\
$NH_3(^1A_1)$     &    -55.447595     &   -55.447969 &-55.449684 &-55.457770 &  -55.468587 &-55.509590\\
$H_2O(^1A_1)$     &    -74.962992     &   -74.965180 &-74.976319 &-75.012500 &   -         &-\\
 \hline
 \end{tabular}
 }
\end{table}

\begin{table}[ht]
{\bf TABLE II}

{\bf Results of ground state test calculations: Percent of the correlation 
energy accounted.}

\centering {
\begin{tabular}{ c c c c c c }
\hline
  & \multicolumn{5}{ c } {$[E_{HF}-E_{\varkappa}]/[{E_{HF}-E_{FCI}}]\times 100\%$} \\
\cline{2-6}
Species &$\varkappa=2$  &$\varkappa=3$    &$\varkappa=4$      &$\varkappa=5$ &$\varkappa=6$ \\
\hline

$Be(^1S)$         &28.7    &42.1&100.0&-        &-  \\
$B(^2P) $         &8.8    &50.0&32.2&100.0    &-\\
$LiH(^1\Sigma ^+)$&11.9    &29.7&100.0& -      &-       \\
$NH_2(^2B_2)$     &1.0    &11.2&34.0& 100.0   &- \\
$NH_3(^1A_1)$     &0.6    &3.4 &16.4&33.9      &100  \\
$H_2O(^1A_1)$     &4.4    &26.9&100&-           &- \\
 \hline
 \end{tabular}
 }
\end{table}

 \end{document}